\documentclass[useAMS,usenatbib]{mn2e}

\usepackage{graphicx}
\usepackage{amsmath,esint}
\usepackage{amssymb}
\usepackage{natbib}
\usepackage{aas_macros}
\usepackage{color}
\usepackage{float}

\usepackage{hyperref}
\hypersetup{
    colorlinks,%
    citecolor=blue,%
    linkcolor=blue,%
    urlcolor=blue
}

\title[Particle acceleration in axisymmetric pulsar current sheets]{Particle acceleration in axisymmetric pulsar current sheets}
\author[Beno\^it Cerutti et al.]{Beno\^it Cerutti\thanks{E-mail:
bcerutti@astro.princeton.edu}\thanks{Lyman Spitzer Jr. Fellow.}, Alexander Philippov, Kyle Parfrey and Anatoly Spitkovsky\\
Department of Astrophysical Sciences, Princeton University, Princeton, NJ 08544, USA.}
\begin{document}

\date{Accepted --. Received --; in original form --}

\pagerange{\pageref{firstpage}--\pageref{lastpage}} \pubyear{2014}

\maketitle

\label{firstpage}

\begin{abstract}
The equatorial current sheet in pulsar magnetospheres is often regarded as an ideal site for particle acceleration via relativistic reconnection. Using 2D spherical particle-in-cell simulations, we investigate particle acceleration in the axisymmetric pulsar magnetosphere as a function of the injected plasma multiplicity and magnetization. We observe a clear transition from a highly charge-separated magnetosphere for low plasma injection with little current and spin-down power, to a nearly force-free solution for high plasma multiplicity characterized by a prominent equatorial current sheet and high spin-down power. We find significant magnetic dissipation in the current sheet, up to $30\%$ within 5 light-cylinder radii in the high-multiplicity regime. The simulations unambiguously demonstrate that the dissipated Poynting flux is efficiently channeled to the particles in the sheet, close to the Y-point within about 1-2 light cylinder radii from the star. The mean particle energy in the sheet is given by the upstream plasma magnetization at the light cylinder. The study of particle orbits shows that all energetic particles originate from the boundary layer between the open and the closed field lines. Energetic positrons always stream outward, while high-energy electrons precipitate back towards the star through the sheet and along the separatrices, which may result in auroral-like emission. Our results suggest that the current sheet and the separatrices may be the main source of high-energy radiation in young pulsars.
\end{abstract}


\section{Introduction}

Young pulsars represent the majority of the high-energy gamma-ray sources identified in our Galaxy \citep{2013ApJS..208...17A}. In spite of exquisite data, the origin of the gamma-ray emission remains poorly understood. The high-energy gamma rays are often associated with curvature radiation, emitted by relativistic electron-positron pairs accelerated in tiny regions where the electric field is not perfectly screened by the plasma (the so-called ``gap'' models, e.g., \citealt{1983ApJ...266..215A, 1986ApJ...300..500C, 1996ApJ...470..469R, 2003ApJ...588..430M}). In this framework, particle acceleration and emission are confined within the corotating magnetosphere delimited by the light cylinder of radius $R_{\rm LC}=Pc/2\pi$, where $P$ is the spin period of the star and $c$ is the speed of light.

There has been significant progress in the numerical modeling of pulsar magnetospheres, mostly in the magnetohydrodynamic (MHD) limit (force-free, resistive force-free, and full MHD, see e.g., \citealt{1999ApJ...511..351C, 2005PhRvL..94b1101G, 2006MNRAS.367...19K, 2006ApJ...648L..51S, 2006MNRAS.368L..30M, 2006MNRAS.368.1055T, 2009A&A...496..495K, 2010ApJ...715.1282B, 2012ApJ...749....2K, 2012MNRAS.424..605P, 2012MNRAS.423.1416P, 2012ApJ...746...60L, 2013MNRAS.435L...1T}), and most recently using particle-in-cell (PIC) simulations \citep{2014ApJ...785L..33P, 2014ApJ...795L..22C}. A prominent feature found in these simulations is the presence of strong current sheets in the magnetosphere. One current sheet forms near the star in each hemisphere at the boundary between the open and the closed field lines (or separatrix). Both sheets merge at the end of the closed zone near the light-cylinder radius (at the ``Y-point''), and create a single current layer which supports open magnetic field lines. \citet{1990ApJ...349..538C} proposed that dissipation in the current sheet via relativistic reconnection could account for the efficient transfer of magnetic flux into energetic particles (i.e., the ``sigma-problem''). \citet{1996A&A...311..172L} pointed out that the observed pulsed gamma-ray emission could be a natural outcome of such a process. In this scenario, particles are accelerated in the current sheet and radiate synchrotron gamma-ray photons \citep{2002A&A...388L..29K, 2012MNRAS.424.2023P, 2013A&A...550A.101A, 2014ApJ...780....3U}.

PIC simulations of isolated current sheets show that relativistic reconnection in collisionless pair plasmas is fast and efficient at accelerating particles (e.g., \citealt{2001ApJ...562L..63Z, 2004PhPl...11.1151J, 2014ApJ...782..104C, 2014ApJ...783L..21S, 2014PhRvL.113o5005G, 2014arXiv1409.8262W}). However, there are important differences between these local simulations and the global structure of the current sheet in pulsars (e.g., geometrical effects, gradients, strong electric field induced by the rotation of the star, absorption/creation of particles), so the results might not be directly applicable. The first kinetic simulations of the aligned pulsar show that the current layers are indeed involved in particle acceleration \citep{2014ApJ...785L..33P, 2014ApJ...795L..22C}, but the details of the acceleration mechanism remains unclear.\\

In this paper, we present a comprehensive analysis of particle acceleration in the aligned pulsar magnetosphere using large, high-resolution two-dimensional (2D) axisymmetric PIC simulations. In addition, this work explores the effect of the plasma supply and magnetization on the global structure of the magnetosphere, and on the pulsar spin-down power. This article is organized as follows. In Sect.~\ref{setup} we describe the implementation of the 2D spherical axisymmetric grid in the {\tt Zeltron} code. Most of the technical details are given in the appendices (Appendix~\ref{app_maxwell}, \ref{app_deposit}). We describe how the simulation is set up, with a particular emphasis on the boundary conditions. In Sect.~\ref{results} we present the main results of this study, which are summarized in Sect.~\ref{conclusions}.

\section{Numerical tools and setup}\label{setup}

\subsection{The {\tt Zeltron} code in spherical coordinates}

In this study, we use the explicit, massively parallel, electromagnetic PIC code {\tt Zeltron}, initially developed for relativistic reconnection studies \citep{2013ApJ...770..147C, 2014ApJ...782..104C}. The code employs the standard second-order accurate Yee algorithm \citep{1966ITAP...14..302Y} to advance the electromagnetic field in time, and the second-order accurate Boris \citep{1991ppcs.book.....B} or \citet{2008PhPl...15e6701V} algorithms to solve the equation of motion of the particles. {\tt Zeltron} does not follow a strictly charge-conserving scheme, instead small errors accumulated in the electric field are corrected periodically by solving Poisson's equation \citep{2003LNP...615....1P}. To perform simulations of the aligned pulsar, we implemented a 2D axisymmetric spherical grid in {\tt Zeltron} (see also the recent developments by \citealt{2014ApJ...795L..22C} and \citealt{2015NewA...36...37B}). Maxwell's equations are discretized and solved on the spherical Yee mesh, using the formulae given in Appendix~\ref{app_maxwell}, while the particles' equations of motion are solved in Cartesian coordinates. The positions and velocities of the particles are then remapped onto the spherical grid where their charges and currents are deposited using the volume weighting technique (bilinear interpolation in $r^3$ and $\cos\theta$) given in Appendix~\ref{app_deposit}. As discussed below, using spherical coordinates greatly simplifies the formulation of the boundary conditions.

\begin{figure}
\centering
\includegraphics[width=8cm]{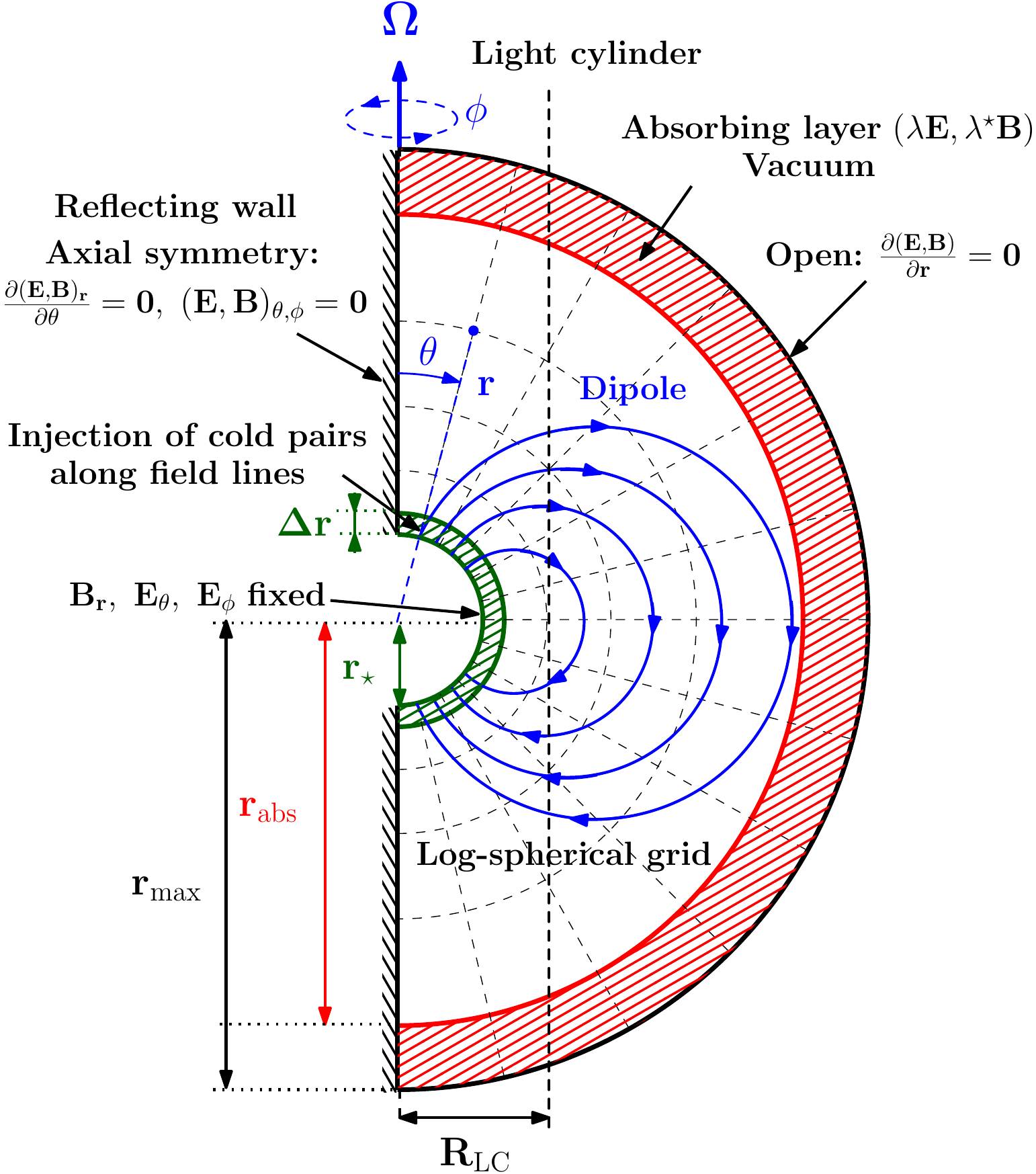}
\caption{Numerical setup and boundary conditions adopted in this study. The initial configuration is composed of a dipole in vacuum anchored on a perfectly conducting star, whose surface coincides with the inner radius of the domain ($r=r_{\star}$), rotating at the angular velocity $\Omega$. Cold pairs are injected in the first row of cells (between $r$ and $r+\Delta r$) along the field lines. A damping layer between $r_{\rm abs}$ and $r_{\rm max}$ absorbs the outgoing electromagnetic waves and plasma. The grid cells are uniformly spaced in $\log_{10}r$ and $\theta$.}
\label{fig_setup}
\end{figure}

\subsection{The grid}

The computational domain extends in radius from the surface of the star, i.e., $r=r_{\star}$, up to $r=r_{\rm max}=6.67 R_{\rm LC}$, and from $\theta=0^{\circ}$ to $\theta=180^{\circ}$ (see Fig.~\ref{fig_setup}). The light-cylinder radius is set at $R_{\rm LC}=3r_{\star}$. The spin axis of the star is aligned along $\theta=0^{\circ}$. The grid points are uniformly spaced in $\log_{10}r$ and $\theta$. The log-spherical grid presents three advantages here: (1) the regions close to the star where the fields are most intense have the highest resolution, (2) the box can be radially extended to large distances at reduced computational costs, and (3) it preserves the cells' aspect ratio. We found that the parallelization of the code for this problem is most optimized if the domain is decomposed along the $\theta$-direction, because the vast majority of particles drift along quasi-radial field lines.

\subsection{The fields}

The simulations start in vacuum with a magnetic dipole frozen into the surface of a perfectly conducting neutron star, and aligned along the rotation axis ($\theta=0^{\circ}$, Fig.~\ref{fig_setup}), such that $\mathbf{\Omega}\cdot\mathbf{B}>0$, where $\Omega$ is the angular velocity of the star. Hence, the three components of the magnetic field are
\begin{eqnarray}
B_{\rm r} &=& \frac{B_{\star}r^3_{\star}\cos\theta}{r^3}\\
B_{\theta} &=& \frac{B_{\star}r^3_{\star}\sin\theta}{2r^3}\\
B_{\phi} &=& 0.
\label{eq_dipole}
\end{eqnarray}
At $t=0$, the star is instantly spun up to its angular velocity $\Omega$ by imposing, on the surface of the star, the poloidal electric field induced by the rotation of the field lines
\begin{eqnarray}
\mathbf{E_{\theta}}( r_{\star},\theta) &=& -\frac{(\mathbf{\Omega}\times\mathbf{R})\times\mathbf{B_{\rm r}}}{c} \nonumber \\
&=& -\frac{r_{\star}\sin\theta}{R_{\rm LC}}B_{\rm r}(r_{\star},\theta)\mathbf{e_{\theta}},
\label{eq_Efield}
\end{eqnarray}
where $R=r\sin\theta$ is the cylindrical radius, and where $\mathbf{e_{\theta}}$ is the unit vector in the $\theta$-directions. The toroidal electric field at the surface of the star is set to $\mathbf{E_{\phi}}(r_{\star},\theta)=\mathbf{0}$ at all times.

The choice of the outer boundary condition for the fields is more involved. To mimic an open boundary in which no information is able to come back inward, we define an absorbing layer starting at $r=r_{\rm abs}$ and extending to the end of the box, where Maxwell's equations contain an electric and a magnetic conductivity terms (see red shaded region in Fig.~\ref{fig_setup}), such that \citep{1991ppcs.book.....B}
\begin{eqnarray}
\frac{\partial \bf E}{\partial t}&=&-\lambda(r)\mathbf{E}+c\left({\bf \nabla}\times{\bf B}\right)-4\pi\mathbf{J}\\
\frac{\partial \bf B}{\partial t}&=&-\lambda^{\star}(r)\mathbf{B}-c\left({\bf \nabla}\times{\bf E}\right).
\end{eqnarray}
One can avoid undesirable reflections of waves at $r=r_{\rm abs}$ by gradually increasing the conductivities with distance. Empirically, we found that the following conductivity profile is a good damping layer,
\begin{equation}
\lambda(r)=\lambda^{\star}(r)=\frac{K_{\rm abs}}{\Delta t}\left(\frac{r-r_{\rm abs}}{r_{\rm max}-r_{\rm abs}}\right)^3, \hspace{0.2cm} {\rm if}~r\geq r_{\rm abs}
\end{equation}
\begin{equation}
\hspace{-3.2cm}{\lambda(r)=\lambda^{\star}(r)=0,} \hspace{0.2cm} {\rm otherwise},
\end{equation}
where $\Delta t$ is the time step (see its definition in Appendix~\ref{app_maxwell}), and $K_{\rm abs}>1$ is a numerical parameter that controls the damping strength. Here, we choose $K_{\rm abs}=40$. At $r=r_{\rm max}$, we apply a zero gradient condition on the fields, i.e., $\partial \mathbf{E}/\partial r=0$ and $\partial \mathbf{B}/\partial r=0$. The absorbing layer is set at $r_{\rm abs}=0.9 r_{\rm max}$ in all the simulations presented here.

Thanks to the integral form of Maxwell's equations derived in Appendix~\ref{app_maxwell}, we are able to push the $\theta$-boundaries to the axis (there is no division by $\sin\theta$). Then one can simply apply the axial symmetry to the fields at $\theta=0$ and $\theta=\pi$: $E_{\theta,\phi}=0,~B_{\theta,\phi}=0$ and $\partial E_{\rm r}/\partial\theta=0,~\partial B_{\rm r}/\partial\theta=0$.

\subsection{The particles}\label{sect_inj}

In principle, one should model the full electromagnetic cascade to obtain a self-consistent injection of pairs into the magnetosphere (e.g., \citealt{2013MNRAS.429...20T, 2014ApJ...795L..22C}). Instead of solving for the cascade, we propose a simple and robust way to fill the magnetosphere with plasma. This method consists of injecting a neutral plasma of pairs at every time step uniformly between the surface of the neutron star $r_{\star}$ and $ r_{\star}+\Delta r$, where $\Delta r$ is the thickness of the first row of cell along the $r$-direction, and with no angular dependence. The particles are injected along the field lines with a poloidal velocity, $v_{\rm pol}=0.5 c$, and in corotation with the star, i.e., with $v_{\phi}=R\Omega$. The flux of particles injected at every time step is
\begin{equation}
F_{\rm inj}=v_{\rm pol}f_{\rm inj} n^{\star}_{\rm GJ},
\label{ninj}
\end{equation}
where $n^{\star}_{\rm GJ}=B_{\star}/(2\pi R_{\rm LC} e)$ is the Goldreich-Julian density \citep{1969ApJ...157..869G} at the pole of the star. This configuration mimics the injection of fresh plasma by the cascade everywhere close to the star. The efficiency of the cascade is parametrized by the injection rate $f_{\rm inj}$. Even though the injection rate of particles per time step is fixed, the plasma density at the surface of the star is free to vary with time and $\theta$, depending on the amount of plasma trapped near the surface and the number of particles returning back to the star. However, to avoid over-injecting into the closed zone where the plasma is trapped, the code adds new pairs only if the plasma at the surface of the star is well magnetized. Plasma magnetization is quantified by the $\sigma$-parameter, defined here as
\begin{equation}
\sigma\equiv\frac{B^2}{4\pi\Gamma(n^{+}+n^{-})m_{\rm e} c^2}
\end{equation}
where $(n^{+}+n^{-})$ is the total plasma density, and $\Gamma$ is the plasma bulk Lorentz factor. We define a minimum value for $\sigma$ at the surface of the star, $\sigma_{\star}\equiv\sigma(r_{\star})\gg 1$, below which no pairs are injected. 

To model the special case of a ``dead'' pulsar where there is no pair production, i.e., the ``electrosphere'' or the ``disk-dome'' solution (see review by \citealt{1999PhR...318..227M}), we used a slightly different method. In this case, the amount of new injected charges is controlled by the surface charge density, $\Sigma$, given by the jump condition
\begin{equation}
4\pi\Sigma=E_{\rm r}(r_{\star})-E^{\rm co}_{\rm r}(r_{\star}),
\end{equation}
where $E_{\rm r}(r_{\star})$ is the radial electric field at the surface of the star, while $E^{\rm co}_{\rm r}(r_{\star})=-(\mathbf{\Omega}\times\mathbf{R})\times\mathbf{B_{\rm \theta}}/c$ is the radial corotating electric field at $r=r_{\star}$. If $\Sigma<0$ ($\Sigma>0$), only electrons (positrons) are injected with the appropriate density, and without any initial velocity along the field lines, i.e., $v_{\rm pol}=0$. In practice, we release only a fraction of $\Sigma$ (a few percent) every time step, to avoid over-injection. The code stops injecting new particles when the density of charge reaches the Goldreich-Julian charge density, $\rho_{\rm GJ}= -(\mathbf{\Omega} \cdot\mathbf{B})/2\pi c$.

Regardless of the injection method, any particle that strikes the $r=r_{\star}$ boundary is removed from the simulation. The same fate applies to any particle leaving the working domain into the absorbing layer, i.e., if $r\geq r_{\rm abs}$. Our results are unchanged if the particles are removed further out, at $r=r_{\rm max}$. On the axis, particles are reflected with no loss of energy.

\subsection{Physical and numerical parameters}

\begin{figure}
\centering
\includegraphics[width=9cm]{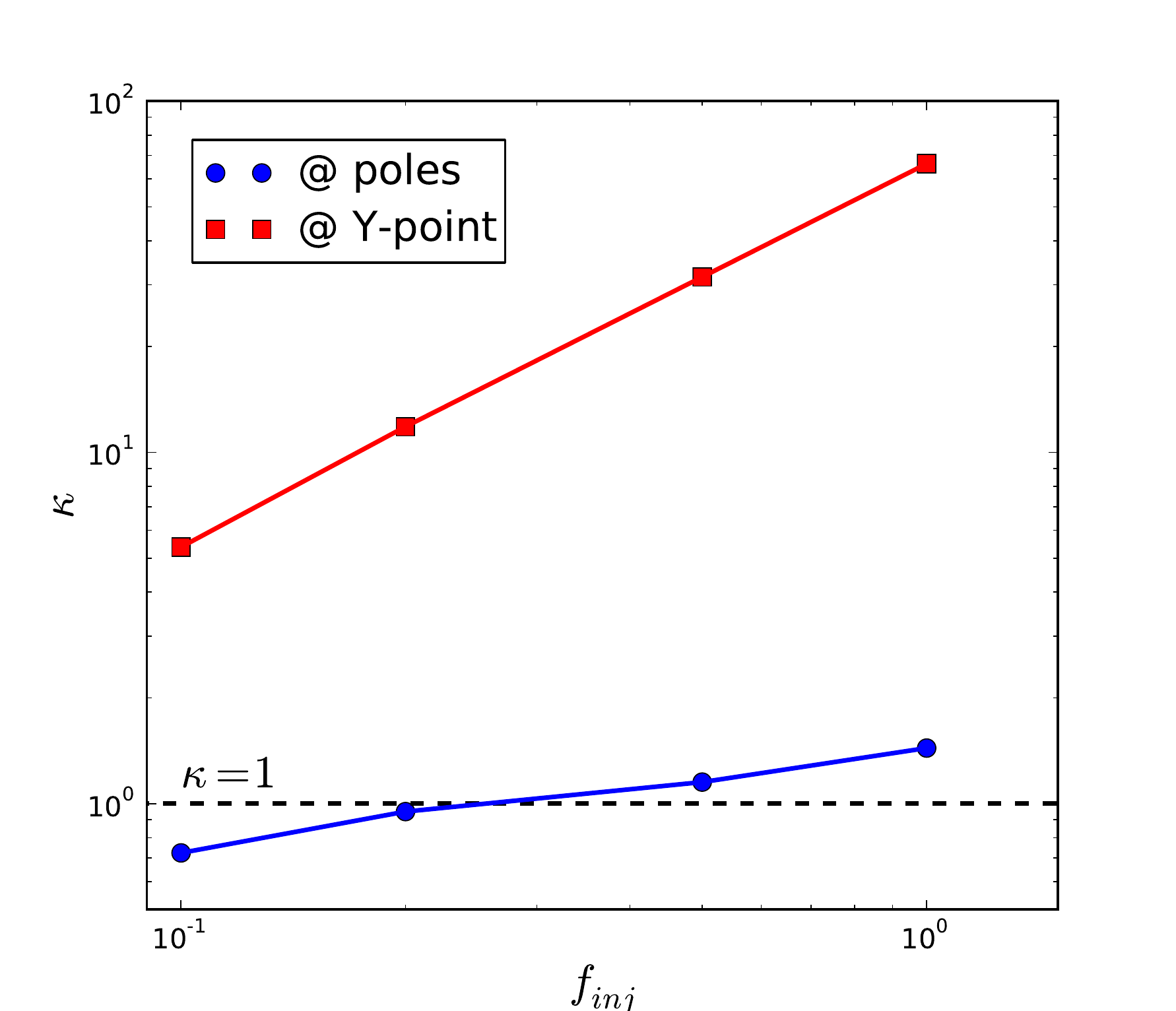}
\caption{Plasma multiplicity ($\kappa=n/n_{\rm GJ}$, where $n_{\rm GJ}$ is the local Goldreich-Julian density) measured in the simulations at the poles (blue dots) and at the Y-point (red squares), as a function of the injection rate $f_{\rm inj}$. The horizontal dashed line shows the critical plasma multiplicity $\kappa=1$.}
\label{fig_multiplicity}
\end{figure}

We perform a series of simulations in which we vary the plasma injection rate $f_{\rm inj}$, and the minimum plasma magnetization parameter defined at the star's surface, $\sigma_{\star}$. In the first set we keep $\sigma_{\star}=5000$ fixed, and study the transition between a charge-starved magnetosphere for $f_{\rm inj}\ll 1$ where pair production is inefficient, to a force-free magnetosphere filled with a dense plasma for $f_{\rm inj}=1$ where $\mathbf{E}\cdot\mathbf{B}=0$ everywhere (except in the current sheet). The force-free regime is appropriate to describe young gamma-ray pulsars which are characterized by intense surface magnetic fields, $B_{\star}\sim 10^{12}~$G, and very large plasma multiplicities $\kappa\gg 1$. The multiplicity is defined as the ratio between the plasma density and the local Goldreich-Julian plasma density $n_{\rm GJ}=|\mathbf{\Omega}\cdot\mathbf{B}|/2\pi e c$. Fig.~\ref{fig_multiplicity} shows the actual local plasma multiplicities achieved in each simulation at the poles and at the Y-point, as a function of $f_{\rm inj}$. The multiplicity increases approximatively linearly with $f_{\rm inj}$. Notice the significant difference in multiplicity between the poles and the Y-point. The multiplicity also increases linearly with the initial kick velocity given to the particles, here fixed at $v_{\rm pol}=0.5 c$. We reproduce the disk-dome solution using the prescription given in Sect.~\ref{sect_inj}. In a second set of simulations, the plasma injection rate is fixed at $f_{\rm inj}=1$ and $\sigma_{\star}$ varies from $500,~1000,~2000$ to $5000$. The purpose of this exploration is to study the effect of the magnetization on particle acceleration in the magnetosphere (Sect.~\ref{sect_acc}).

In real pulsars, there is a huge separation of scales between the plasma skin-depth, $d_{\rm e}$, the radius of the star and the light-cylinder radius, which cannot be reproduced in the simulations. However, it is possible to scale the problem down by decreasing the gap between these scales, and still keep the same hierarchy of scales, i.e., $d_{\rm e}\ll r_{\star}<R_{\rm LC}$. In the runs with $f_{\rm inj}=1$, we have $d_{\rm e}/r_{\star}\approx 1.5\times 10^{-3}$ and $r_{\star}/R_{\rm LC}=0.33$. Our fiducial simulation is composed of $1024$ cells in $r$ and in $\theta$, which gives a spatial resolution at the surface of the star of $d_{\rm e}/\Delta r\approx 5$ for $f_{\rm inj}=0.1$ down to $d_{\rm e}/\Delta r\approx 0.5$ for $f_{\rm inj}=1$. The plasma frequency is well resolved in all runs, with $\Delta t\leq 0.28\omega_{\rm p}^{-1}$. Note that these estimates apply for non-relativistic particles, with a Lorentz factor $\gamma=1$. In reality $\gamma\gg 1$ everywhere in the magnetosphere; hence all plasma scales are very well resolved. Each simulation ran for about 4 spin periods, which is sufficient to establish a quasi-steady state everywhere, even in the outer parts of the box. The number of particles per cell injected at every time step (at the inner boundary) is set at 8 (4 per species). Note that, once the magnetosphere is established the actual number of particles per cell is typically of order 50. We checked that our results are numerically converged by increasing the number of grid cells to $2048^2$, and by varying the number of injected particles per boundary cell from 2 to 32.

\section{results}\label{results}

\subsection{Establishment of the magnetosphere}

\begin{figure*}
\centering
\includegraphics[width=8.5cm]{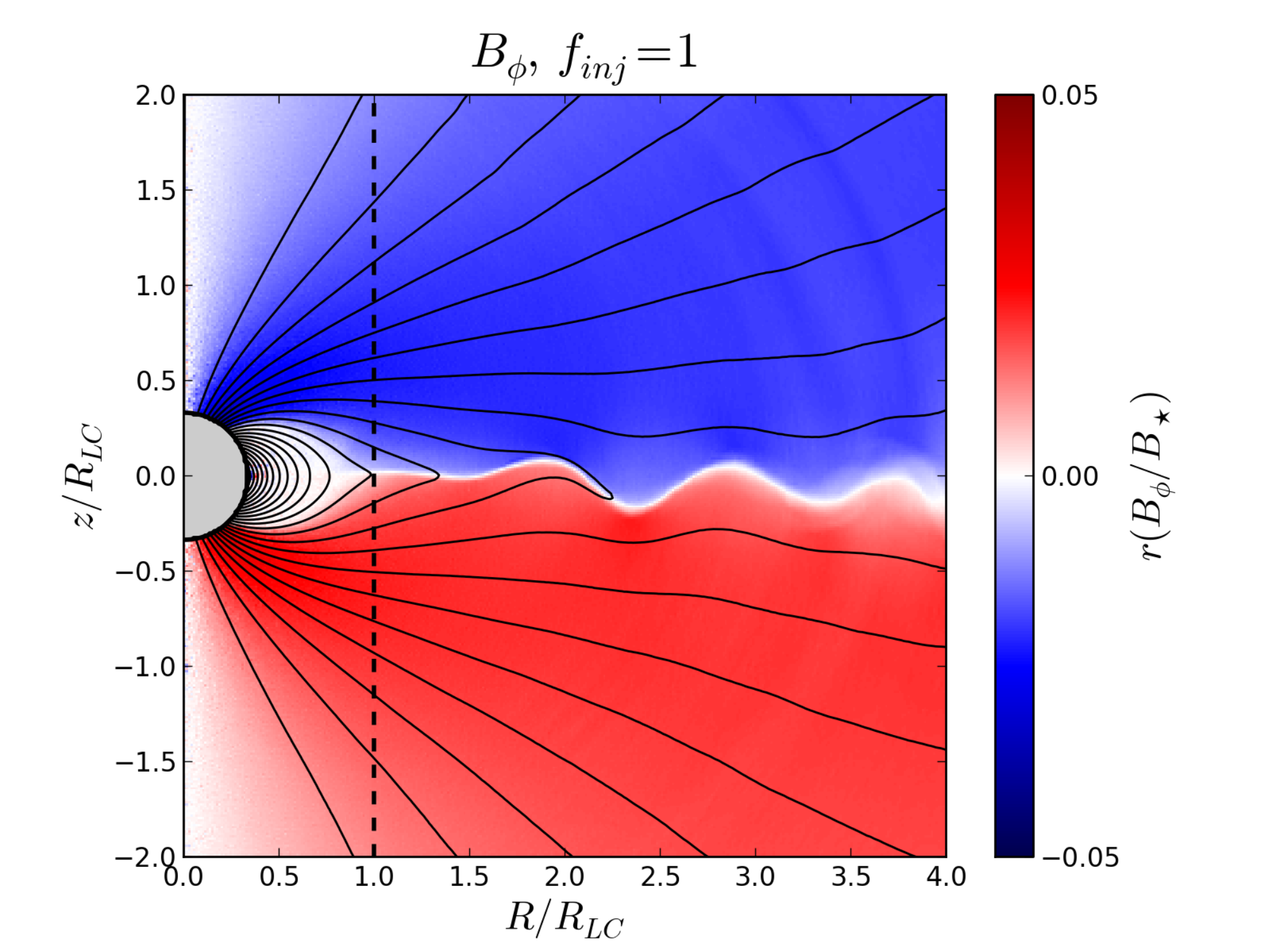}
\includegraphics[width=8.5cm]{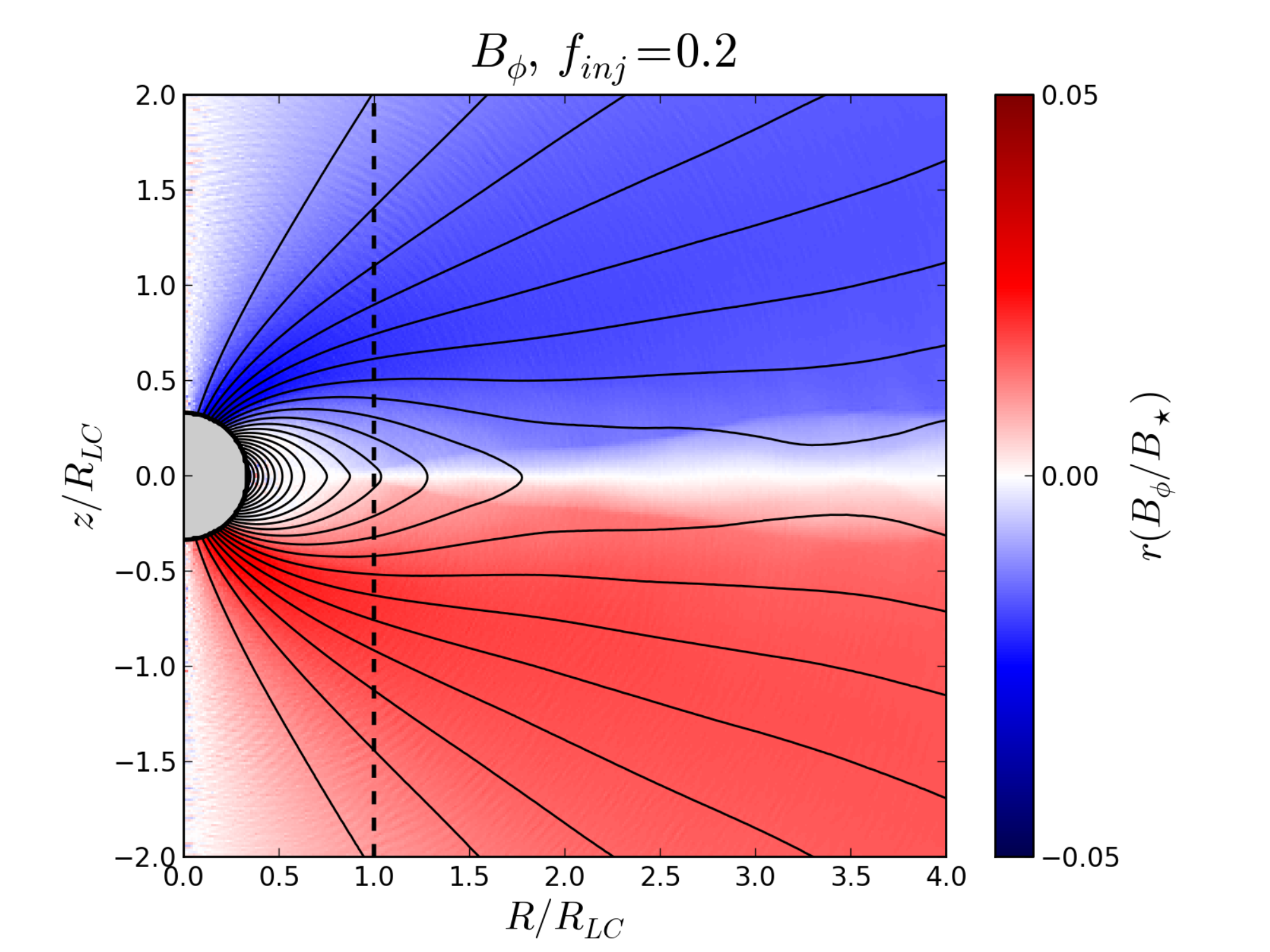}
\includegraphics[width=8.5cm]{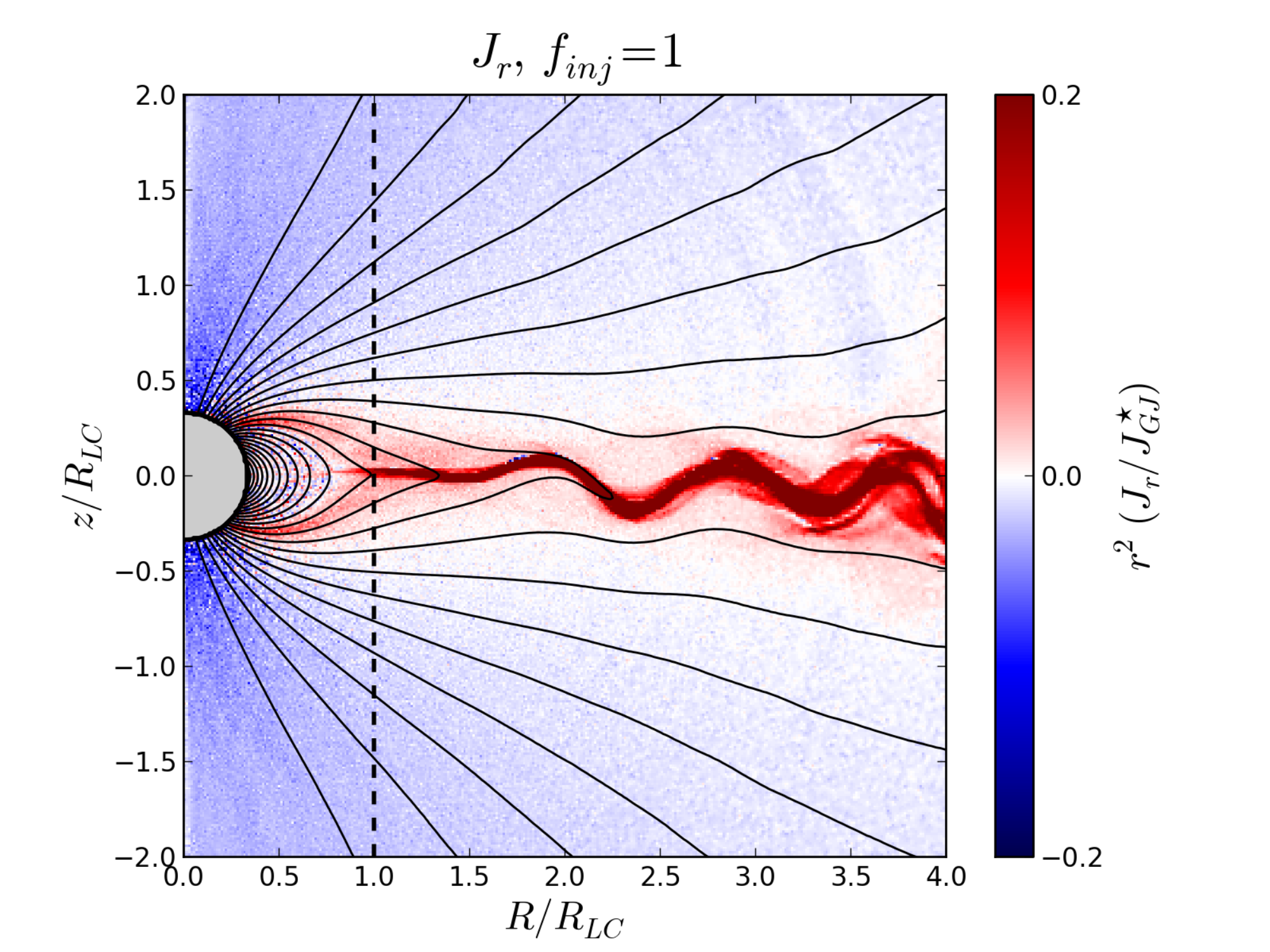}
\includegraphics[width=8.5cm]{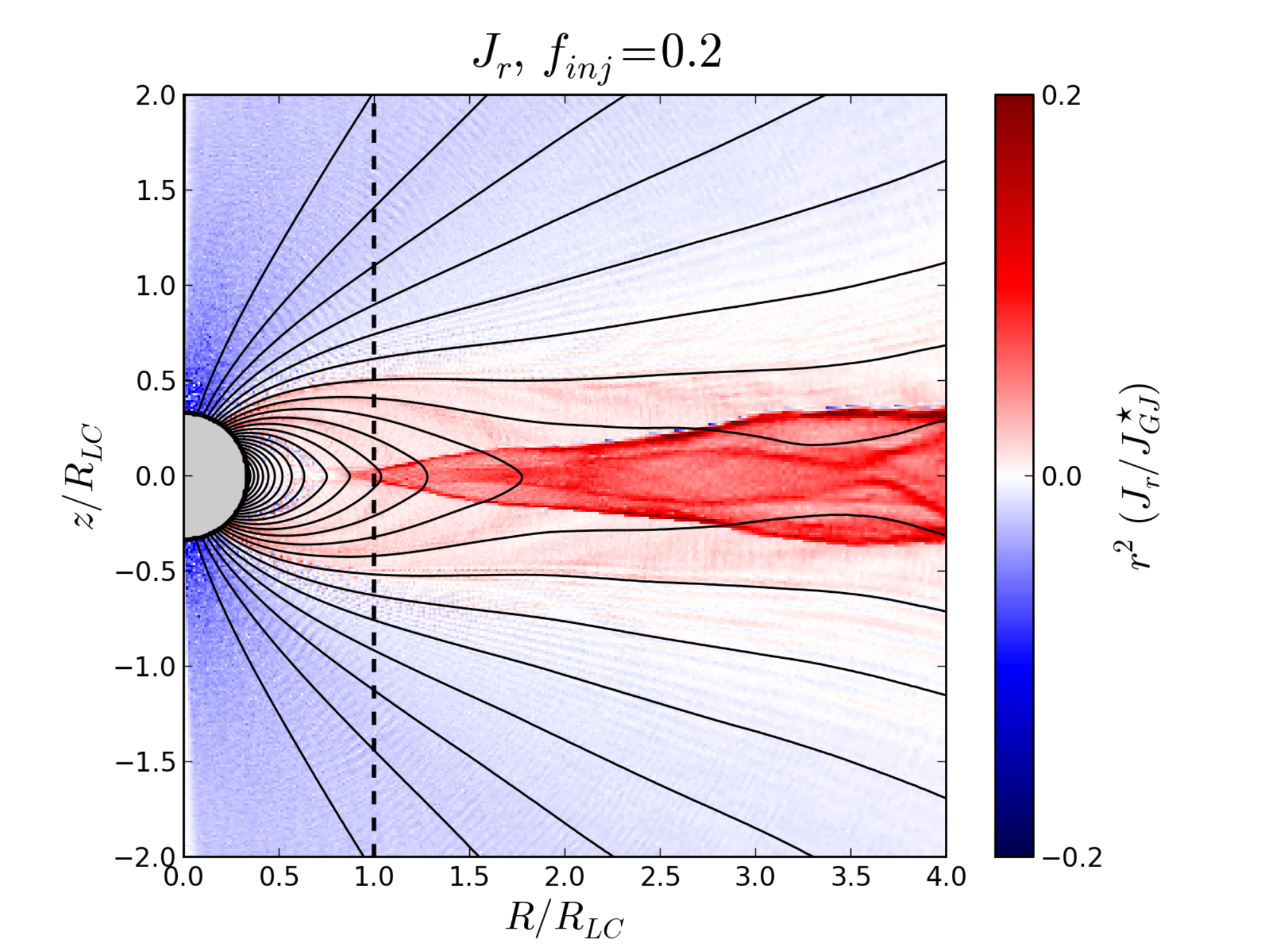}
\includegraphics[width=8.5cm]{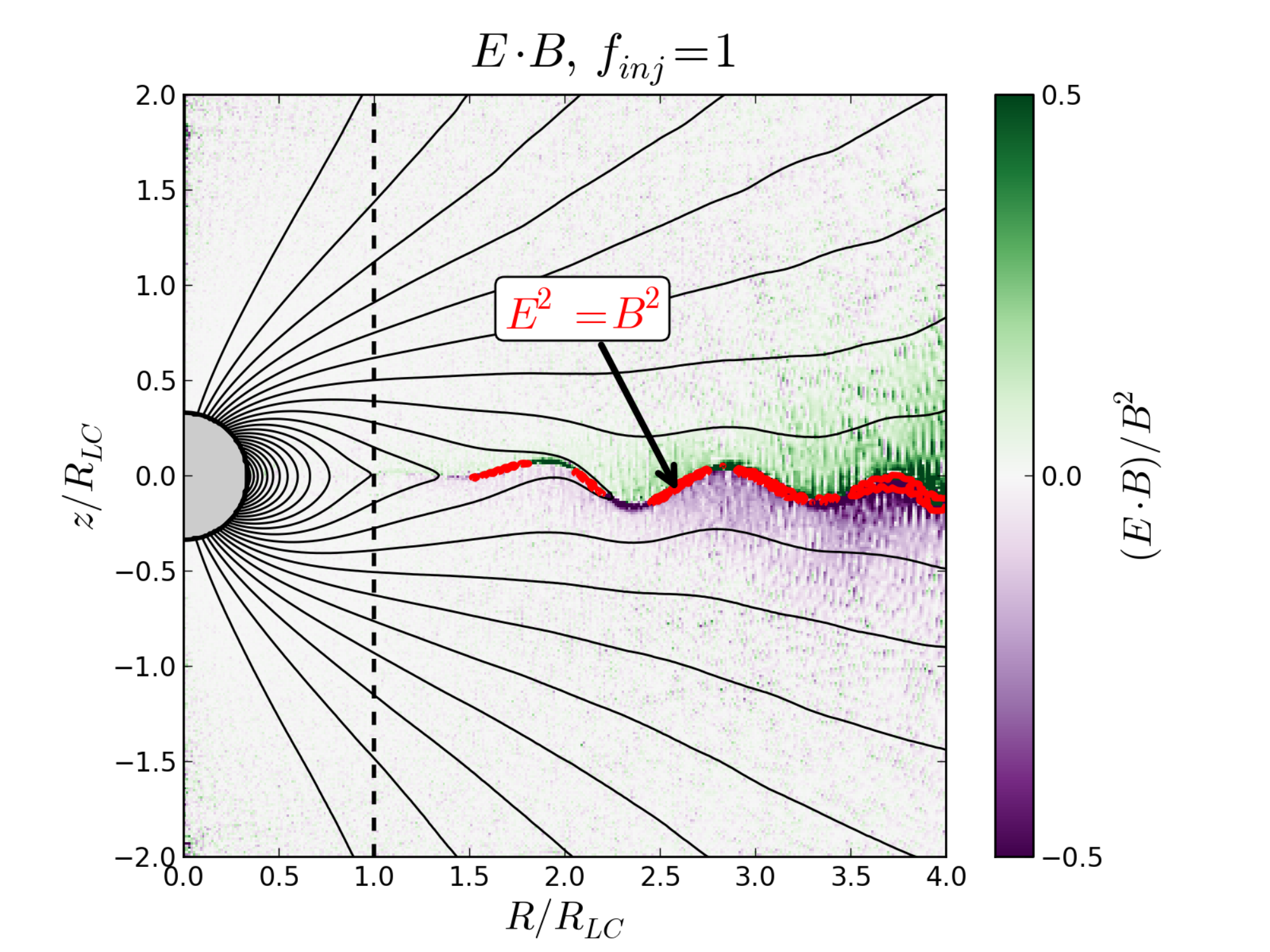}
\includegraphics[width=8.5cm]{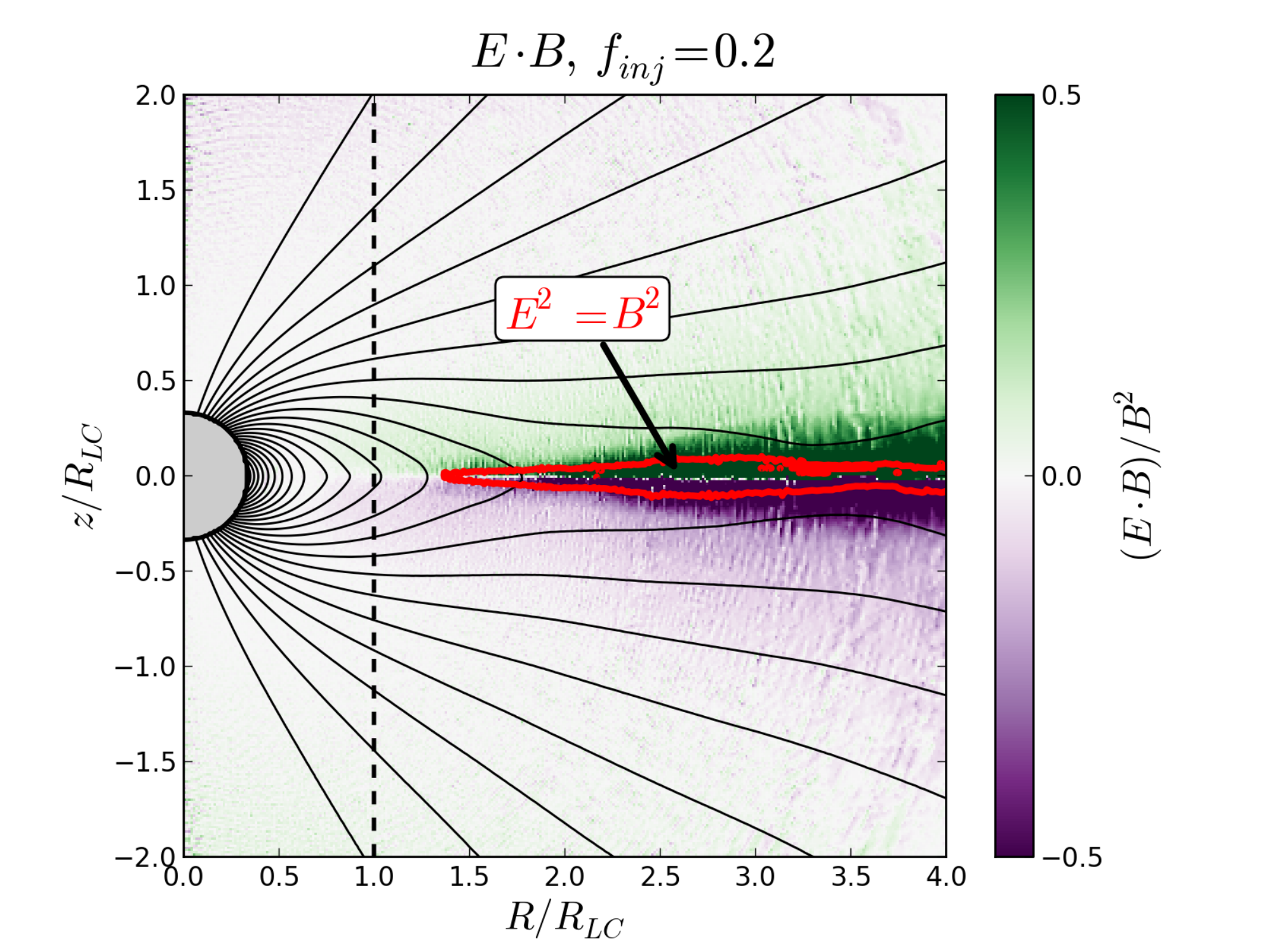}
\caption{Structure of the aligned pulsar magnetosphere, for $f_{\rm inj}=1$ at $t=4P$ (left) and $f_{\rm inj}=0.2$ at $t=3P$ (right). The top panels show the toroidal magnetic field, normalized by the surface magnetic field at the poles, $B_{\star}$, and multiplied by $r$. The middle panels present the radial current density, normalized by the Goldreich-Julian current density at the pole $J^{\star}_{\rm GJ}$, and multiplied by $r^2$. The bottom panels show the relativistic invariant $\mathbf{E}\cdot\mathbf{B}$ normalized by $\mathbf{B}^2$, and the contour where $\mathbf{E}^2=\mathbf{B}^2$ (red solid line) indicated by the black arrows. In all panels, the black solid lines represent the poloidal magnetic field lines, which are given by the isocontours of the magnetic flux function. The vertical dashed line indicates the light-cylinder radius. The gray semi-disk represents the neutron star.} 
\label{fig_magnetosphere}
\end{figure*}

\begin{figure*}
\centering
\includegraphics[width=8.5cm]{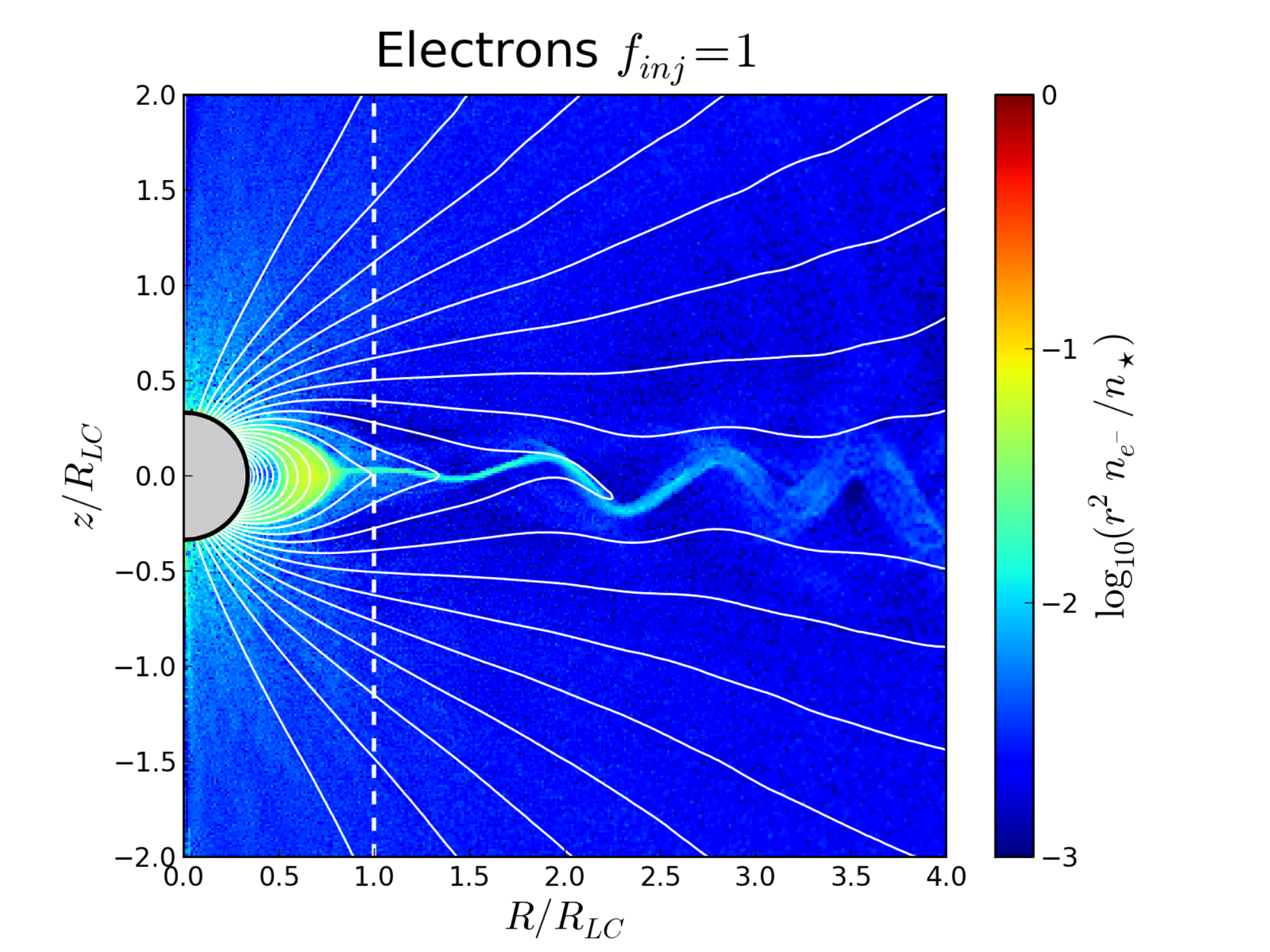}
\includegraphics[width=8.5cm]{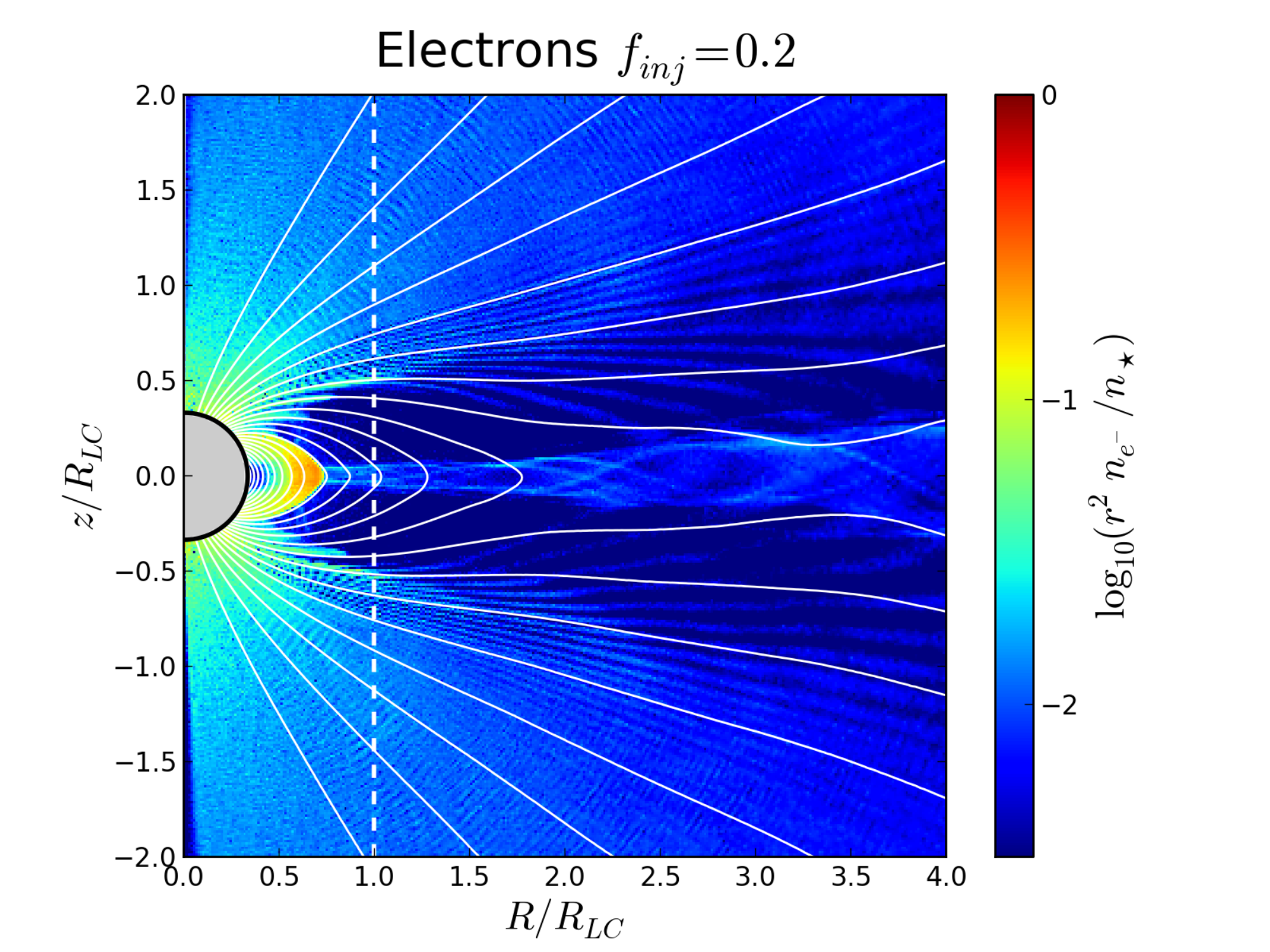}
\includegraphics[width=8.5cm]{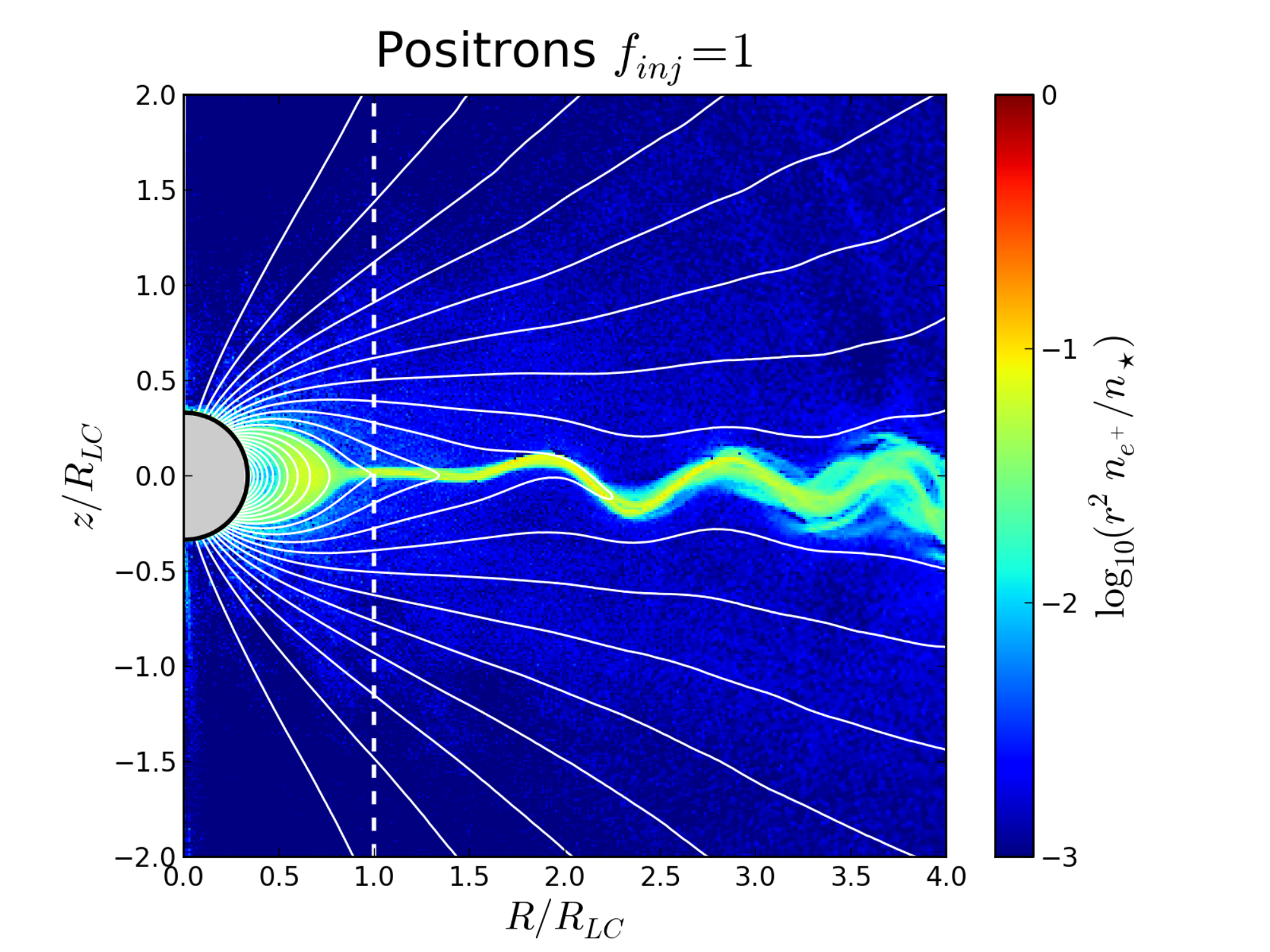}
\includegraphics[width=8.5cm]{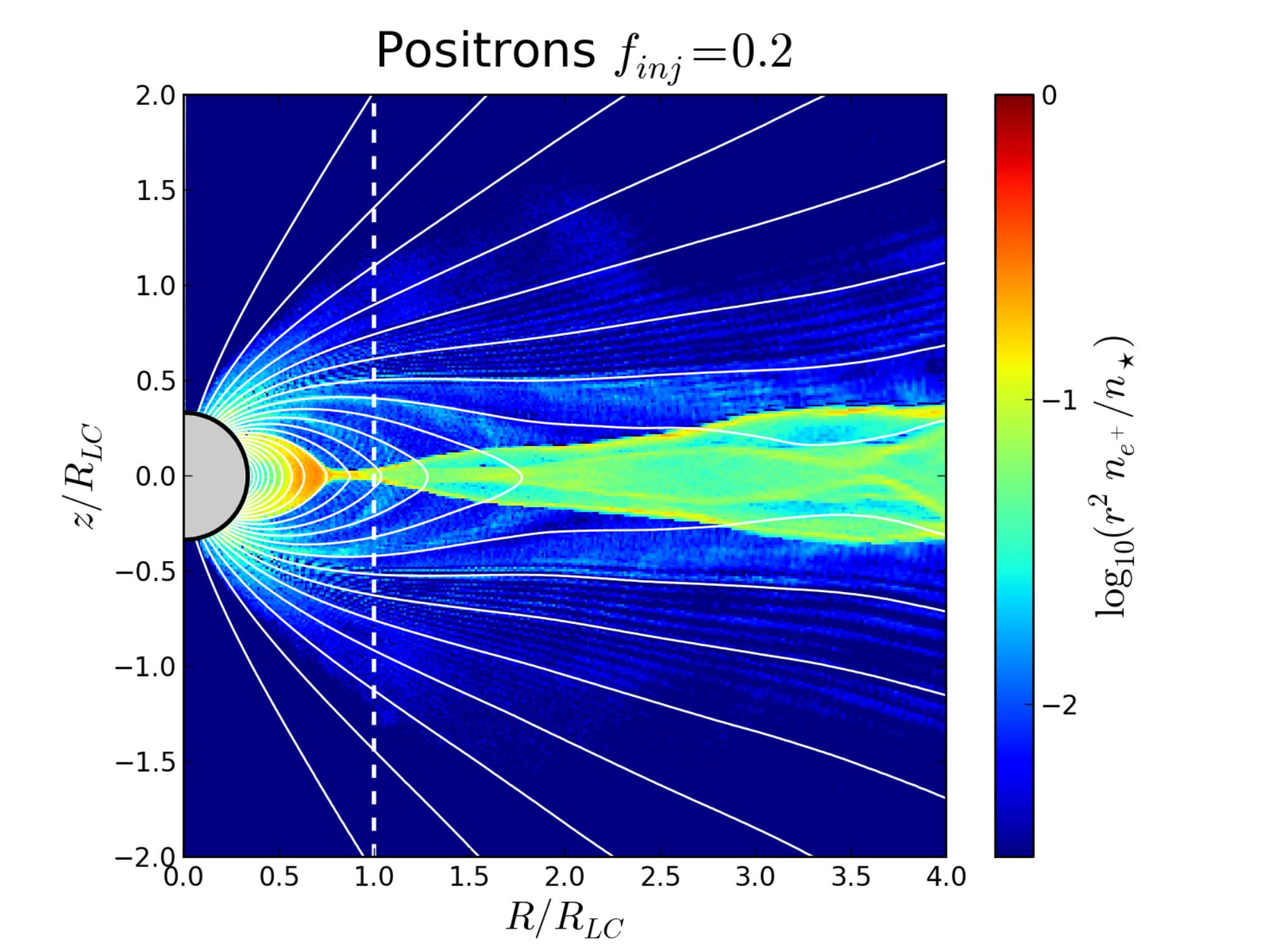}
\caption{Electron (top) and positron (bottom) densities, for $f_{\rm inj}=1$ at $t=4P$ (left), and $f_{\rm inj}=0.2$ at $t=3P$ (right). Densities are normalized to the plasma density at the pole of the star $n_{\star}$, and multiplied by $r^2$. White solid lines are the poloidal magnetic field lines, and the white vertical dashed line is the light-cylinder radius.}
\label{fig_density}
\end{figure*}

All the simulations begin with the launch from the stellar surface of a torsional spherical Alfv\'en wave, that propagates outward at nearly the speed of light. This wave distributes the poloidal electric field throughout the box and sets the magnetosphere in rotation. The magnetosphere settles quickly into a quasi-steady state behind the wave. The initial transient stage ends when the wave is absorbed by the damping layer. Fig.~\ref{fig_magnetosphere} presents (from top to bottom) the structures of the toroidal magnetic field, the radial current densities, the relativistic invariant $\mathbf{E}\cdot\mathbf{B}$ and the regions where $\mathbf{E}^2>\mathbf{B}^2$, for $f_{\rm inj}=1$ at $t=4P$ (left), and $f_{\rm inj}=0.2$ at $t=3P$ (right). Fig.~\ref{fig_density} shows the corresponding electron (top) and positron (bottom) densities.

In the case where the magnetosphere is fed with a large amount of plasma ($f_{\rm inj}=0.5,~1$, and $\kappa> 1$ at the poles, see left column of Figs.~\ref{fig_magnetosphere},~\ref{fig_density}), the solution is close to force-free (e.g., \citealt{1999ApJ...511..351C}) with $\mathbf{E}\cdot\mathbf{B}\approx 0$ almost everywhere, except in the current layers where the force-free approximation breaks down ($\mathbf{E}\cdot\mathbf{B}\neq 0$ and $\mathbf{E}^2>\mathbf{B}^2$, see bottom-left panel in Fig.~\ref{fig_magnetosphere}). The current sheet forms self-consistently in the equator and supports open magnetic field lines beyond the light-cylinder radius. After about one rotation period, and before the flow reaches the absorbing layer, the current sheet becomes unstable and starts flapping around the equator in the $\theta$-direction at $r\gtrsim 2R_{\rm LC}$ (see also \citealt{2014ApJ...785L..33P}). The amplitude of the kink increases roughly linearly with distance as the wind expands. At ${\rm r\gg R_{\rm LC}}$ and away from the current sheet, the magnetic field structure asymptotically approaches the split monopole solution \citep{1973ApJ...180L.133M}, where $B_{\rm r}\propto 1/r^2$ and $B_{\phi}\propto 1/r$. Both the toroidal and the poloidal magnetic fields reverse across the current layer, supported respectively by the radial and the toroidal currents. The current returns back into the star via the polar caps (middle panel in Fig.~\ref{fig_magnetosphere}). The plasma is not perfectly neutral: the polar regions are negatively charged, while the current sheet is positively charged. As shown below, the current is not only sustained by moving charged plasma but also by the counter-streaming of both species, in particular in the current sheets (see Sect.~\ref{sect_bulk}).

The structure of the magnetosphere in the low-density regime ($f_{\rm inj}=0.1,~0.2$, and $\kappa< 1$ at the poles, see right column of Figs.~\ref{fig_magnetosphere},~\ref{fig_density}) changes significantly. The force-free description is not valid anymore ($\mathbf{E}\cdot\mathbf{B}\neq 0$, bottom-right panel in Fig.~\ref{fig_magnetosphere}), and large vacuum gaps (regions with almost no plasma) open between the polar regions and the equator. The magnetosphere is almost perfectly charge-separated, with electrons at the poles and positrons at the equator. Notice that particles cross field lines in the equator (in particular near the light-cylinder), because $E>B$ in this region. There is still an electric current flowing through the magnetosphere, but with lower intensity, which explains why fewer magnetic field lines open. The current layer forms a wedge with an opening angle $\sim 10^{\rm o}$ for $f_{\rm inj}=0.2$ and, in contrast with the high-multiplicity solution, it remains perfectly stable throughout the simulation. In this regime, the current does not form sharps arcs along the separatrices as in the high-multiplicity case. There is no clear Y-point either. Instead, the outflowing current is more diffuse and connect smoothly with the equatorial current layer beyond the light cylinder. This solution resembles the electrostatically supported current layer model recently proposed by \citet{2014ApJ...781...46C}.

\begin{figure}
\centering
\includegraphics[width=8.5cm]{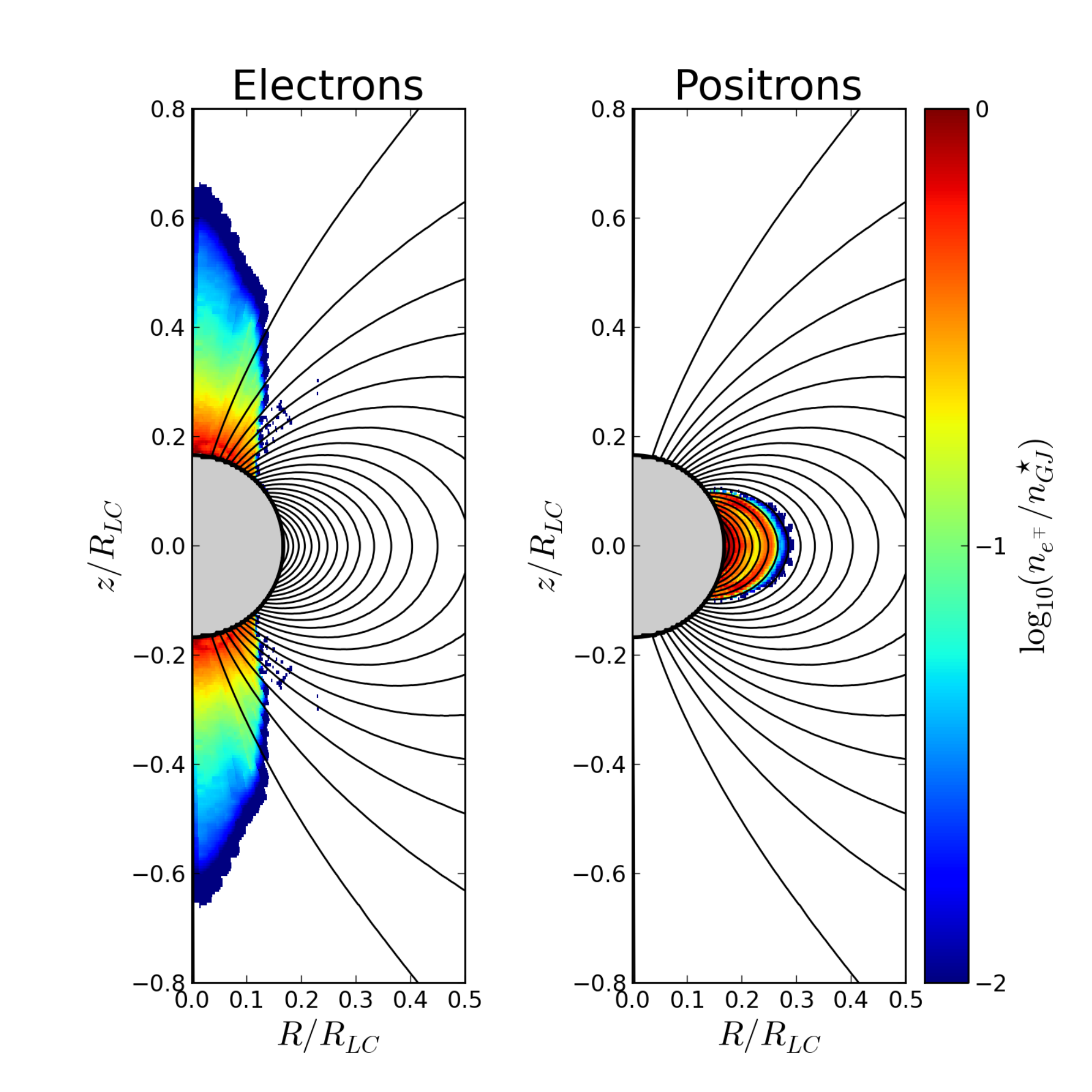}
\caption{Electron (left) and positron (right) densities in the disk-dome solution, obtained after one rotation period. Densities are normalized to the Goldreich-Julian density at the neutron star's pole, $n^{\star}_{\rm GJ}$.}
\label{fig_electrosphere}
\end{figure}

\begin{figure}
\centering
\includegraphics[width=9cm]{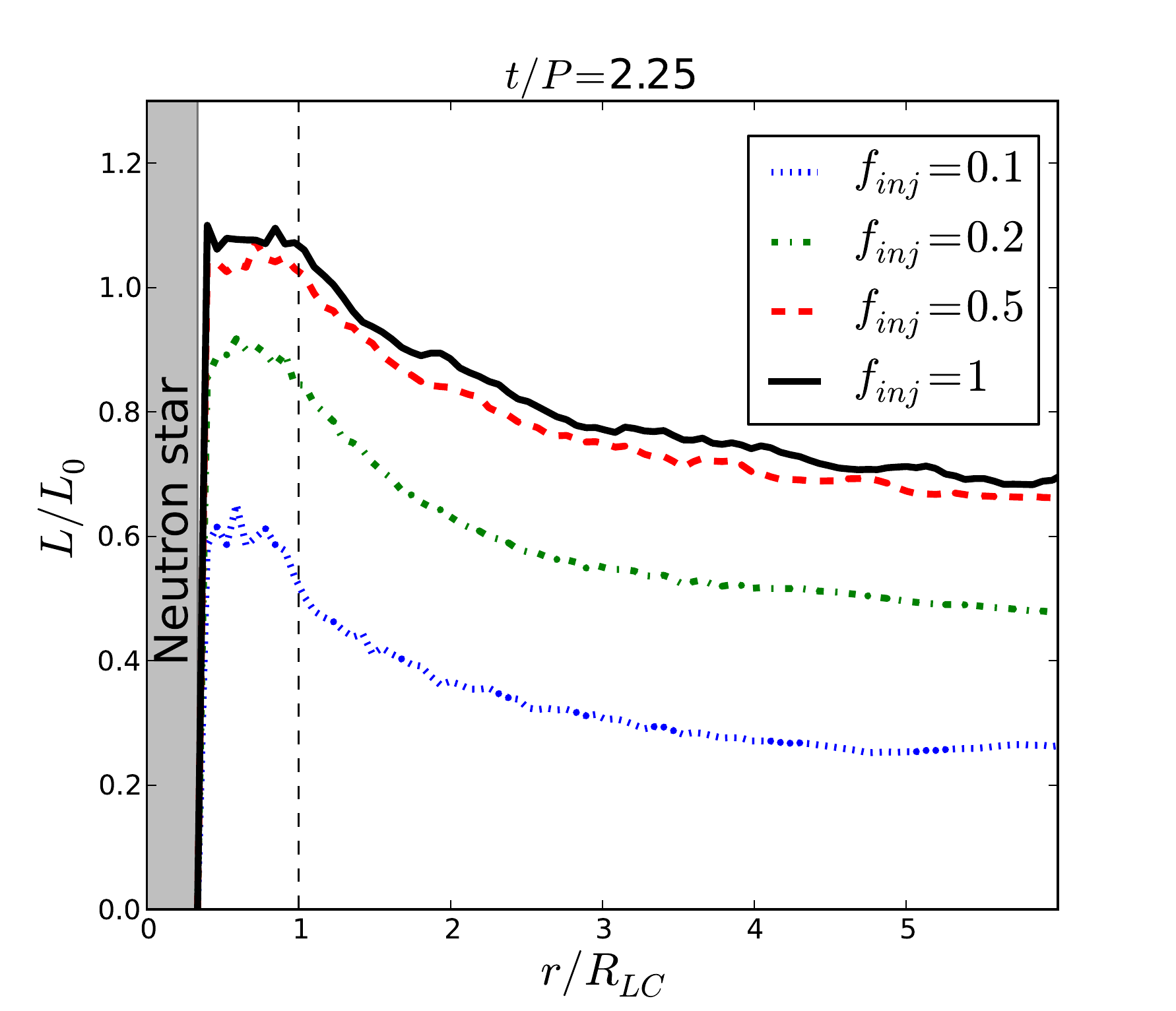}
\caption{Outgoing Poynting flux integrated over a sphere of radius $r$, as function of radius at time $t=2.25P$, for $f_{\rm inj}=0.1$ (blue dotted line),~0.2 (green dashed-dotted line),~0.5 (red dashed line), and 1 (black solid line). The energy flux is normalized to the analytic spin-down power of an aligned pulsar, $L_0=c B_{\star}^2r_{\star}^6/4R^4_{\rm LC}$. The disk-dome solution is not shown here because it does not spin down, $L(r)=0$.}
\label{fig_spindown}
\end{figure}

In the extreme case where there is no pair production, we find that the solution reaches the disk-dome configuration, in agreement with previous studies \citep{1985A&A...144...72K, 2001MNRAS.322..209S, 2002A&A...384..414P, 2002ASPC..271...81S, 2014ApJ...785L..33P}. This solution consists of a purely electrostatic structure with trapped positrons around the equator and trapped electrons at the poles, with no outflows or electric currents (see Fig.~\ref{fig_electrosphere}).

Fig.~\ref{fig_spindown} presents the total outgoing Poynting flux integrated over a sphere of radius $r$, or luminosity, as a function of radius, $L(r)$. The energy flux at the light cylinder increases with the plasma injection rate and saturates for $f_{\rm inj}\geq 0.5$ to the spin-down power of an aligned rotator, i.e., $L(R_{\rm LC})\approx L_0=c B_{\star}^2r_{\star}^6/4R^4_{\rm LC}$, in agreement with previous works in the force-free regime (e.g., \citealt{1999ApJ...511..351C, 2005PhRvL..94b1101G, 2006ApJ...648L..51S, 2012MNRAS.423.1416P, 2014ApJ...785L..33P}). The disk-dome solution does not spin down, as expected. An important result is the large dissipation of the Poynting flux with distance, even in the high-multiplicity solution. At $r=5R_{\rm LC}$, the flux decreases by about $30\%$ for $f_{\rm inj}\geq 0.5$. In fact, half of the dissipation occurs between $r=R_{\rm LC}$ and $r=2R_{\rm LC}$. This result is robust against the number of particles per cell and resolution. In addition, we did not find a clear dependence with the ratio $r_{\star}/R_{\rm LC}$, but we prefer not to draw any conclusion as the range explored here, from $r_{\star}/R_{\rm LC}=0.17$ to $0.5$, is rather modest. \citet{2014ApJ...785L..33P} also reported the same amount of dissipation at $r=2R_{\rm LC}$ (a $15\%$ drop). As shown in Sect.~\ref{sect_acc}, the dissipated energy is efficiently channeled into energetic particles within the sheet. When the plasma supply in the magnetosphere decreases, the dissipation rate increases to about $40\%$ for $f_{\rm inj}=0.2$, and to about $50\%$ for $f_{\rm inj}=0.1$. Interestingly, we notice that the $f_{\rm inj}=0.2$ solution gives the same dissipation rate and spin-down power ($\approx 0.82L_0$) as found by \citet{2014ApJ...781...46C}. It is also close to the spin-down power reported by \citet{2014ApJ...795L..22C} if copious pair creation occurs in the current sheet (their ``type I'' solution).

\subsection{Bulk motions}\label{sect_bulk}

With these simulations, we are able to measure the relative motion between the two species in the magnetosphere and see how this contributes to the total current density. For this purpose, we reconstruct the fluid velocity vector, $\mathbf{V}$, from the motion of individual particles averaged in every grid cell. In relativistic kinetic theory, one defines the fluid velocity vector as
\begin{equation}
\mathbf{V}=\frac{\int\left(\frac{\mathbf p}{\mathcal{E}}\right)f\left(\mathbf{r},\mathbf{p}\right)d^3\mathbf{p}}{\int f\left(\mathbf{r},\mathbf{p}\right)d^3\mathbf{p}},
\end{equation}
where $\mathbf{p}$ is the particle momentum, $\mathcal{E}=\gamma m_{\rm e} c^2$ is the total particle energy, and $f\left(\mathbf{r},\mathbf{p}\right)$ is the particle distribution function.

\begin{figure}
\centering
\includegraphics[width=8.5cm]{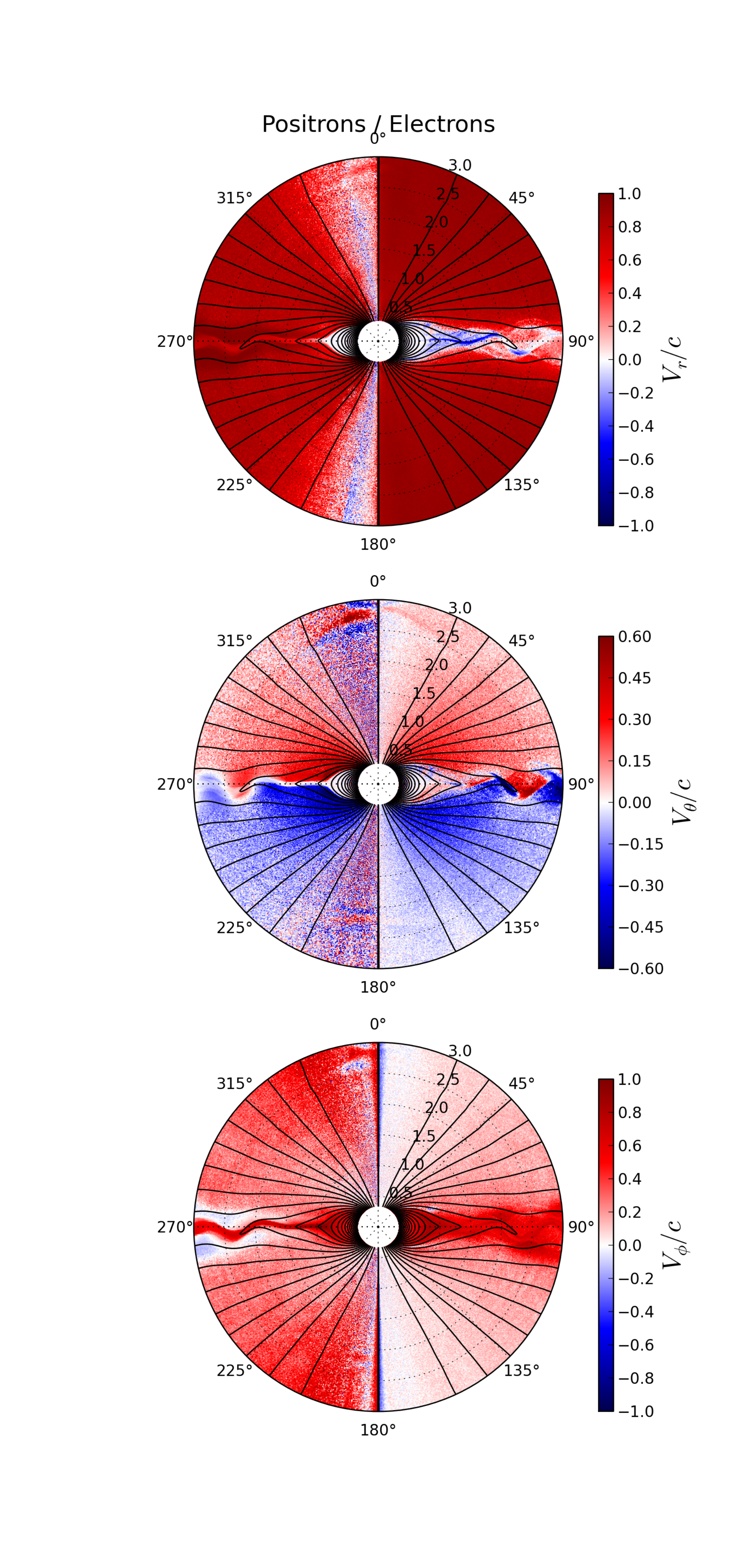}
\caption{Components of the positron (left) and electron (right) fluid velocity map, $V_{\rm r}/c$ (top), $V_{\theta}/c$ (middle), $V_{\phi}/c$ (bottom), at $t=4P$ for $f_{\rm inj}=1$,~$\sigma_{\star}=5000$.}
\label{fig_bulk}
\end{figure}

Fig.~\ref{fig_bulk} shows the fluid velocity maps of each component, for both positrons (left panels) and electrons (right panels) in the inner magnetosphere at time $t=4P$, where $f_{\rm inj}=1$ and $\sigma_{\star}=5000$. In the radial direction (top panel), both the electrons and positrons fly outward along the open field lines almost everywhere with a velocity of order $V_{\rm r}/c\approx +0.8-0.9$ at $r\approx R_{\rm LC}$ and increasing with radius, with the notable exceptions of the polar regions and in the current sheets. This radial motion is due to the $\mathbf{E_{\theta}}\times\mathbf{B_{\phi}}$ drift. At the poles, electrons and positrons both fly away from the star but at different velocities. Electrons move at $V_{\rm r}/c\approx +0.8$ while positrons move at about $\rm V_{\rm r}/c\approx +0.1$ only. Hence, in addition to the net negative charge density at the poles, this difference in velocity between the two species also contributes to the negative radial current shown in Fig.~\ref{fig_magnetosphere}, middle panels.

In the current sheet, we discover that the two species counter-propagate with very high speeds: positrons are ejected outward at $+0.6c$, while electrons precipitate back towards the star at about $-0.6c$ (appearing as a bright blue beam in Fig.~\ref{fig_bulk}, top right panel). As discussed below, this result has important consequences for particle acceleration and radiation. The $V_{\theta}$-component does not show significant differences between electrons and positrons. Both species have mildly relativistic ($\approx 0.5 c$) drift velocities from the open field lines towards the current sheets, the highest speeds being closest to the sheet. These converging flows maintain and feed the current sheets with fresh plasma. They do not appear in the split-monopole model (e.g., \citealt{1973ApJ...180L.133M}), because they are caused by reconnection in the equatorial layer which is captured by the simulations (see also \citealt{2013MNRAS.435L...1T}). The fluid velocity along the $\phi$-direction highlights the fast rotation about the axis of the plasma in the current sheet.

\begin{figure}
\centering
\includegraphics[width=9cm]{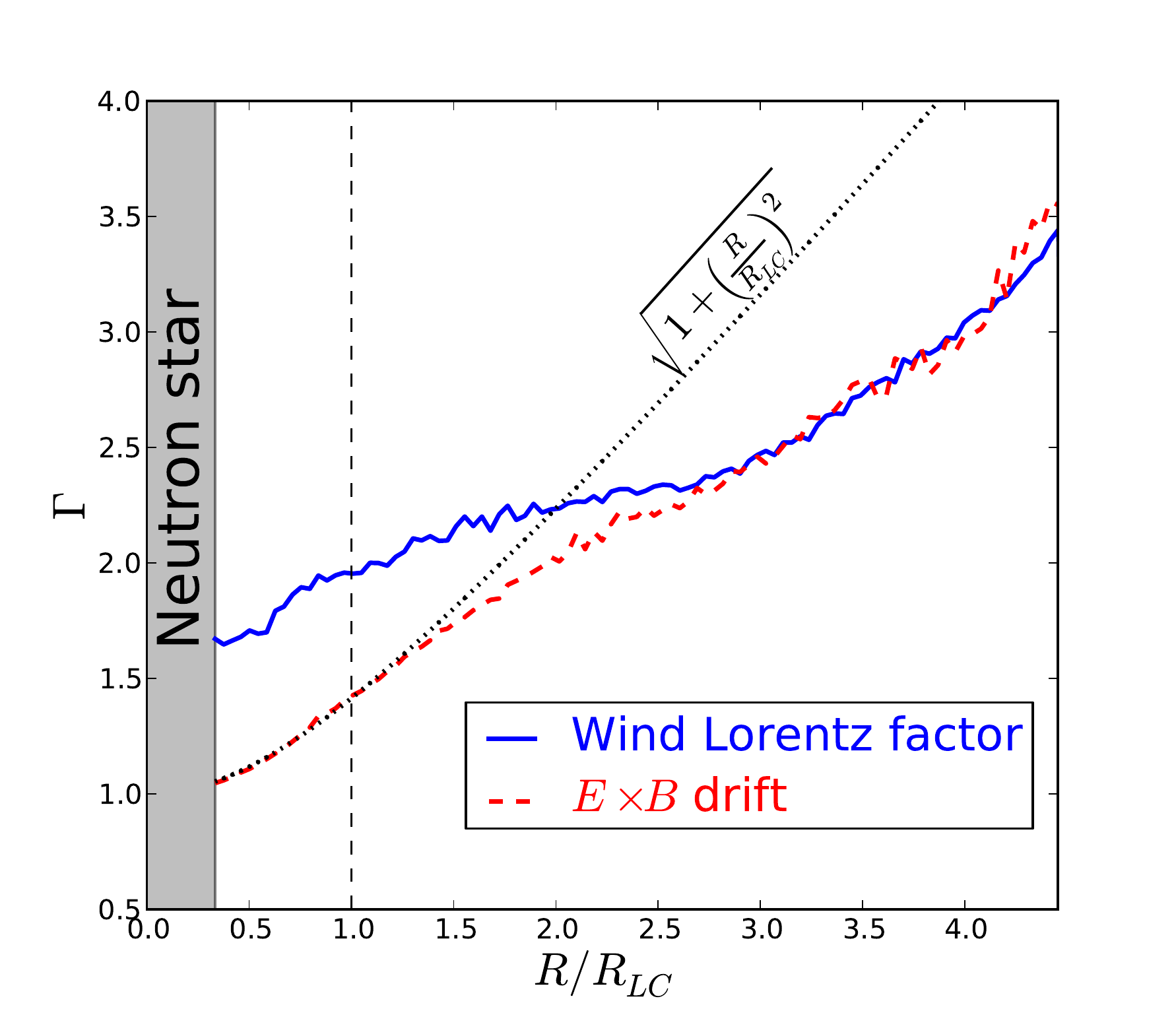}
\caption{Wind Lorentz factor, $\Gamma=1/\sqrt{1-\mathbf{V}^2/c^2}$ (blue solid line), averaged along the $z$-direction ($z=r\cos\theta$) as a function of the cylindrical radius $R=r\sin\theta$. The red dashed line shows the $z$-averaged Lorentz factor of a particle moving exactly at the $\mathbf{E}\times\mathbf{B}$ drift velocity. For comparison, the dotted line shows the analytical solution $\Gamma=\sqrt{1+(R/R_{\rm LC})^2}$ found by \citet{2002ApJ...566..336C}.}
\label{fig_wind}
\end{figure}

By combining all these data together, we infer the bulk Lorentz factor of the wind, $\Gamma=1/\sqrt{1-(\mathbf{V}/c)^2}$. Fig.~\ref{fig_wind} presents the variations of $\Gamma$ as a function of the cylindrical radius ($R=r\sin\theta$), averaged along the $z$-direction (where $z=r\cos\theta$). This figure also compares the Lorentz factor of the wind with the Lorentz factor of a particle moving at the $\mathbf{E}\times\mathbf{B}$ drift velocity, $\Gamma_{\rm drift}$ (red dashed line in Fig.~\ref{fig_wind}). The wind starts with $\Gamma\approx 1.6$ which depends on the initial velocity of the injected particles. Then, $\Gamma$ increases slowly with distance until it coincides with the $\mathbf{E}\times\mathbf{B}$ drift Lorentz factor at $R\approx 2.5 R_{\rm LC}$. At this point, the wind Lorentz factor follows almost exactly the linear increase of $\Gamma_{\rm drift}$ with distance, in agreement with previous works \citep{1977MNRAS.180..125B, 2002ApJ...566..336C, 2006MNRAS.367...19K, 2006MNRAS.368.1717B}. We note that $\Gamma_{\rm drift}$ follows perfectly the $\Gamma_{\rm drift}=\sqrt{1+(R/R_{\rm LC})^2}$ solution found by \citet{2002ApJ...566..336C} within the light cylinder (dotted line in Fig.~\ref{fig_wind}). Our solution deviates from this analytical expression in the region $1\lesssim R/R_{\rm LC}\lesssim 3$, but it tends back to the expected linear increase of $\Gamma$ with $R$ further out. At the outer edge of the box ($R=4.5 R_{\rm LC}$), the wind Lorentz factor reaches about $\Gamma\approx 3.5$. In principle, the wind should accelerate up to the fast magnetosonic point, beyond which the wind Lorentz factor should saturate due to the dominant inertia of the plasma, but our box is too small to observe this transition.

\subsection{Particle acceleration}\label{sect_acc}

\begin{figure}
\centering
\includegraphics[width=9.5cm]{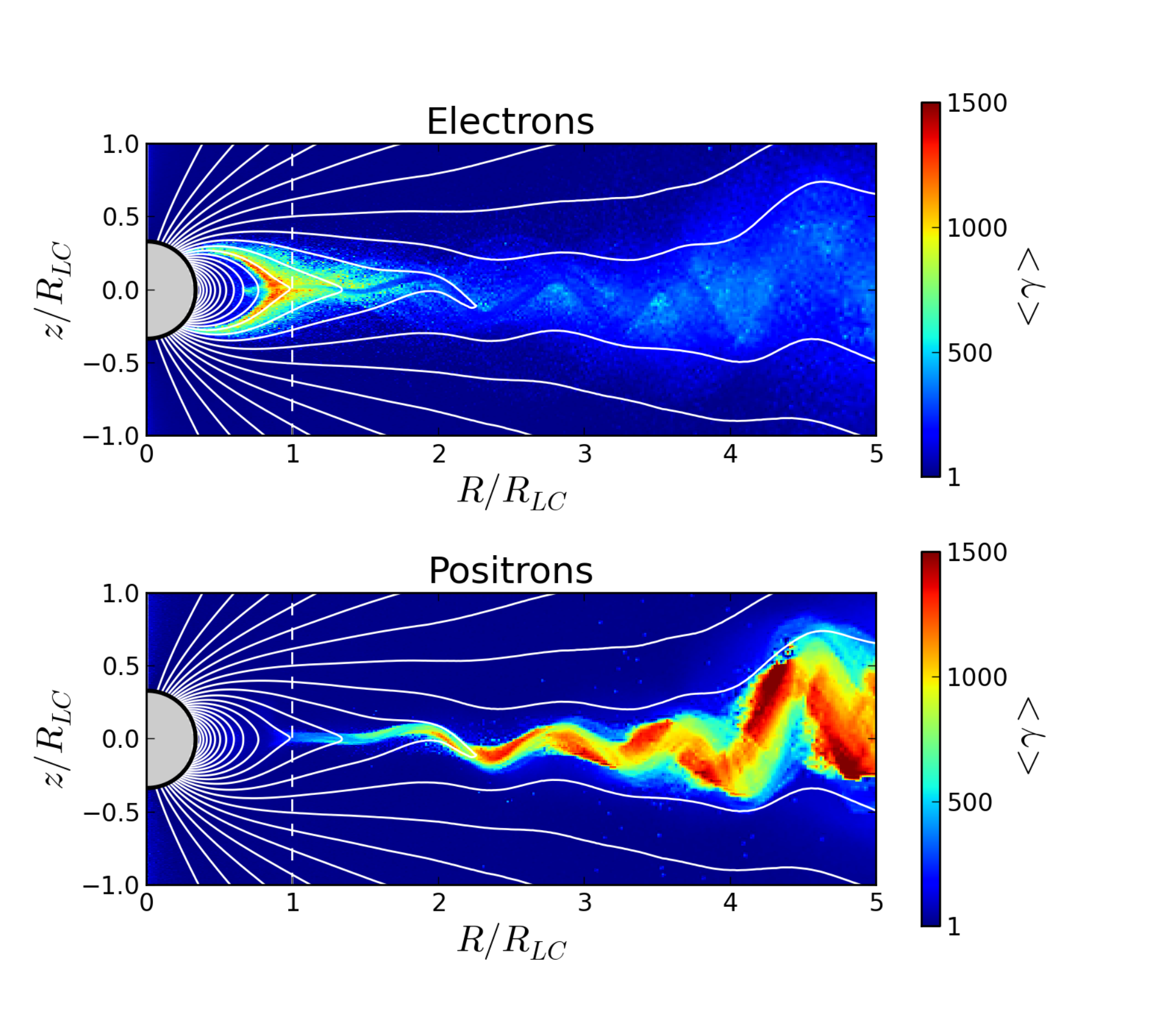}
\caption{Mean electron (top) and positron (bottom) Lorentz factor in the magnetosphere at $t=3.5 P$, with $f_{\rm inj}=1,~\sigma_{\star}=5000$.}
\label{fig_temperature}
\end{figure}

\begin{figure}
\centering
\includegraphics[width=9cm]{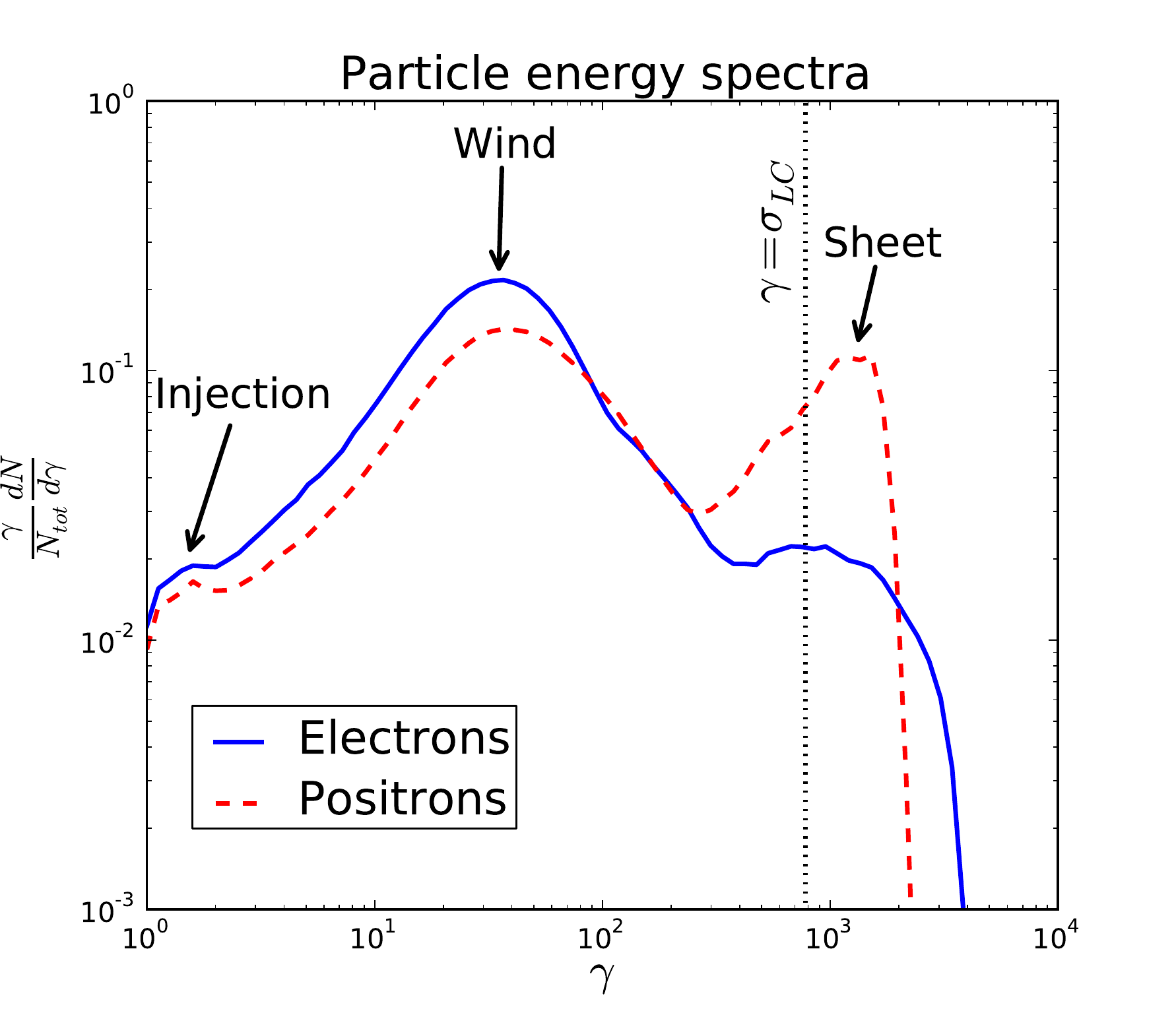}
\caption{Total electron (blue solid line) and positron (red dashed line) energy spectra, $\gamma/N_{\rm tot}(dN/d\gamma)$, for $f_{\rm inj}=1$, $\sigma_{\star}=5000$. The arrows show three characteristic Lorentz factors in the spectra: (i) $\gamma\approx 1.5$ for the injected particles; (ii) $\gamma\approx 30$ for the particles in the wind; (iii) $\gamma\approx \sigma_{\rm LC}=780$ for the particles accelerated in the equatorial current sheet, where $\sigma_{\rm LC}$ is the plasma magnetization at the light cylinder.}
\label{fig_spectra}
\end{figure}

\begin{figure}
\centering
\includegraphics[width=9cm]{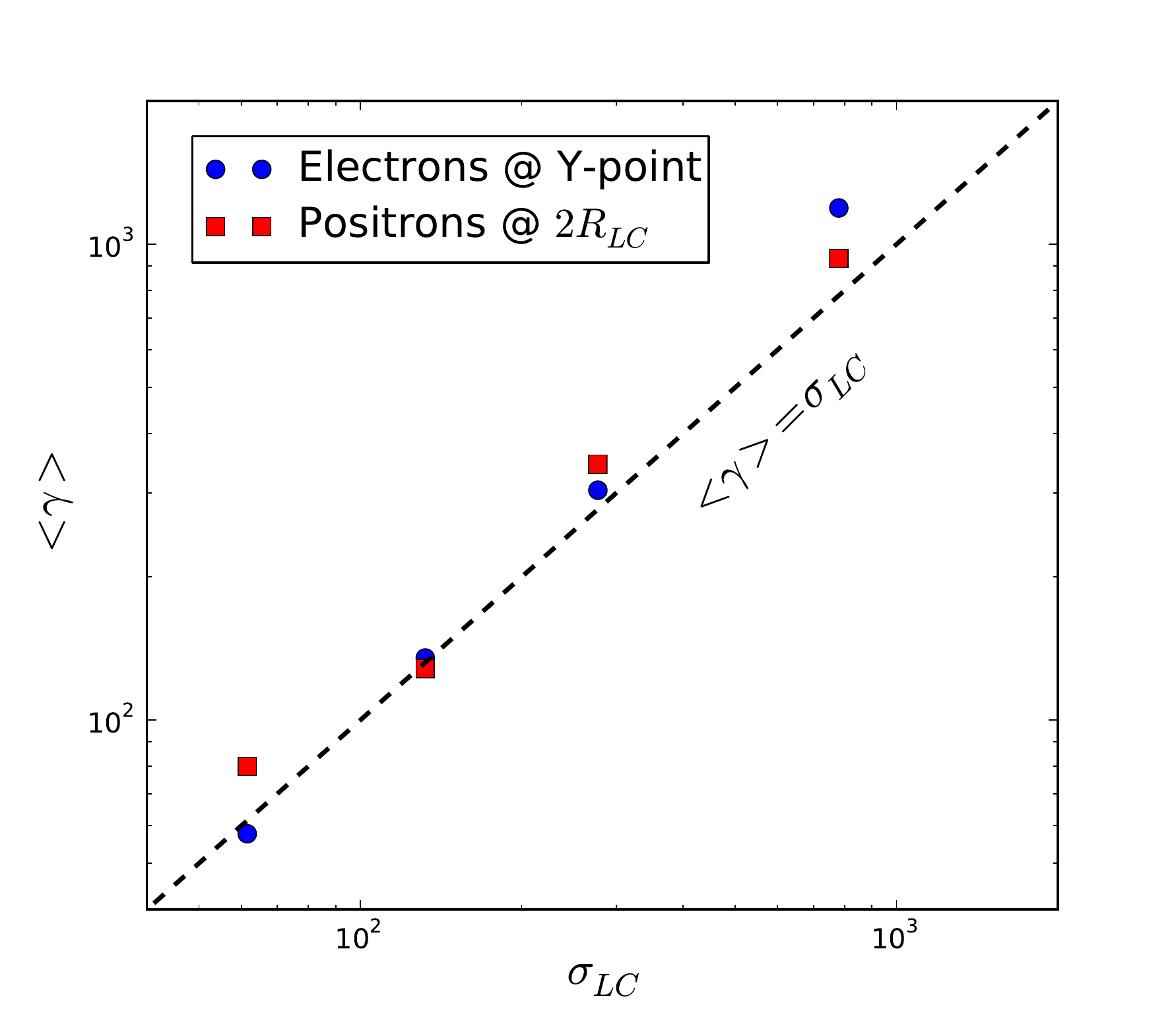}
\caption{Relationship between the magnetization parameter of the plasma at the light cylinder (away from the current layer), $\sigma_{\rm LC}$, and the mean electron (blue disks) and positron (red squares) Lorentz factor $\langle\gamma\rangle$, at $t=3.5 P$. The electrons' mean Lorentz factor is measured at the Y-point ($r=R_{\rm LC},~\theta=90^{\rm o}$), while the positrons' mean Lorentz factor is measured in the current sheet, at $r=2R_{\rm LC}$. $f_{\rm inj}=1$ in all simulations, and $\sigma_{\star}=500,~1000,~2000,$ and 5000, respectively corresponding to $\sigma_{\rm LC}= 62,~132,~277,$ and 780. The dashed line shows $\langle\gamma\rangle=\sigma_{\rm LC}$.}
\label{fig_sigma}
\end{figure}

Fig.~\ref{fig_temperature} presents the mean Lorentz factor for each species averaged over each cell throughout the pulsar magnetosphere at the end of the simulation, for the highest-multiplicity solution ($f_{\rm inj}=1$ and $\sigma_{\star}=5000$). This figure unambiguously demonstrates the key role of the current sheet in the acceleration of particles \citep{2014ApJ...785L..33P, 2014ApJ...795L..22C}, but it also highlights an important difference between electrons and positrons. The high-energy positrons are all located beyond the light-cylinder radius, and are confined within the kinking equatorial sheet. Their mean energy is $\langle\gamma\rangle m_{\rm e}c^2\approx 1000 m_{\rm e}c^2$, which represents about $17\%$ of the total electric potential drop as discussed below. Energetic electrons reach energies similar to those of the positrons in the equatorial sheet, but they concentrate almost exclusively around the Y-point and the current sheets along the separatrices. This suggests that each species follows a very different path before being accelerated (see below).

Fig.~\ref{fig_spectra} shows the energy spectra of all the particles in the box at the end of the simulation. Below $\gamma=200$, the electron and the positron spectra are very similar; they are both composed of a broad bump peaking at $\gamma\approx 30$. These particles are part of the pulsar wind. Above $\gamma=200$, the positron spectrum presents a second sharp peak at $\gamma\approx \langle\gamma\rangle\approx 1000$, as intense as the wind component, followed by an abrupt cut-off. The electron spectrum also extends up to $\gamma\gtrsim 1000$, but the number of electrons is about 10 times smaller than the number of positrons in this band. The reason for this deficit will become clear in the following. We find that the mean energy reached by the particles in the sheet is given by the magnetization parameter at the light cylinder, away from the equatorial layer, i.e,
\begin{equation}
\langle\gamma\rangle\approx \sigma_{\rm LC}\equiv\frac{B^2_{\rm LC}}{4\pi \Gamma_{\rm LC}(n^+_{\rm LC}+n^-_{\rm LC}) m_{\rm e}c^2},
\label{eq_gamma}
\end{equation}
where $B_{\rm LC}$, $\Gamma_{\rm LC}\approx 2$, and $n^+_{\rm LC}\approx n^-_{\rm LC}=\kappa n_{\rm GJ}$ are respectively the magnetic field strength, the wind bulk Lorentz factor, and the positron/electron densities at the light cylinder. This result is expected if the dissipated magnetic energy is transferred to the plasma and shared equally between the particles. It is also possible to interpret this result in terms of the total potential drop across the open field line region, $\Phi_0=B_{\star}r_{\star}^3\Omega^2/c^2$. Using $n_{\rm GJ}\approx \Omega B_{\rm LC}/2\pi e c$ and $B_{\rm LC}\approx B_{\star}(r_{\star}/R_{\rm LC})^3$, one can rewrite Eq.~(\ref{eq_gamma}) as
\begin{equation}
\langle\gamma\rangle\approx\sigma_{\rm LC}\approx\frac{e\Phi_0}{4\Gamma_{\rm LC}\kappa m_{\rm e}c^2}.
\label{eq_phi0}
\end{equation}
For $f_{\rm inj}=1$, the plasma multiplicity in the wind is $\kappa\approx 1.5$ (Fig.~\ref{fig_multiplicity}). Hence, Eq.~(\ref{eq_phi0}) indicates that the particles in the sheet experience on average about 17\% of the full potential drop. Some electrons undergo almost 50\% of $\Phi_0$. Fig.~\ref{fig_sigma} shows that the $\langle\gamma\rangle\approx\sigma_{\rm LC}$ relation applies for all values of $\sigma_{\rm LC}$ explored in this study.

\begin{figure}
\centering
\includegraphics[width=9cm]{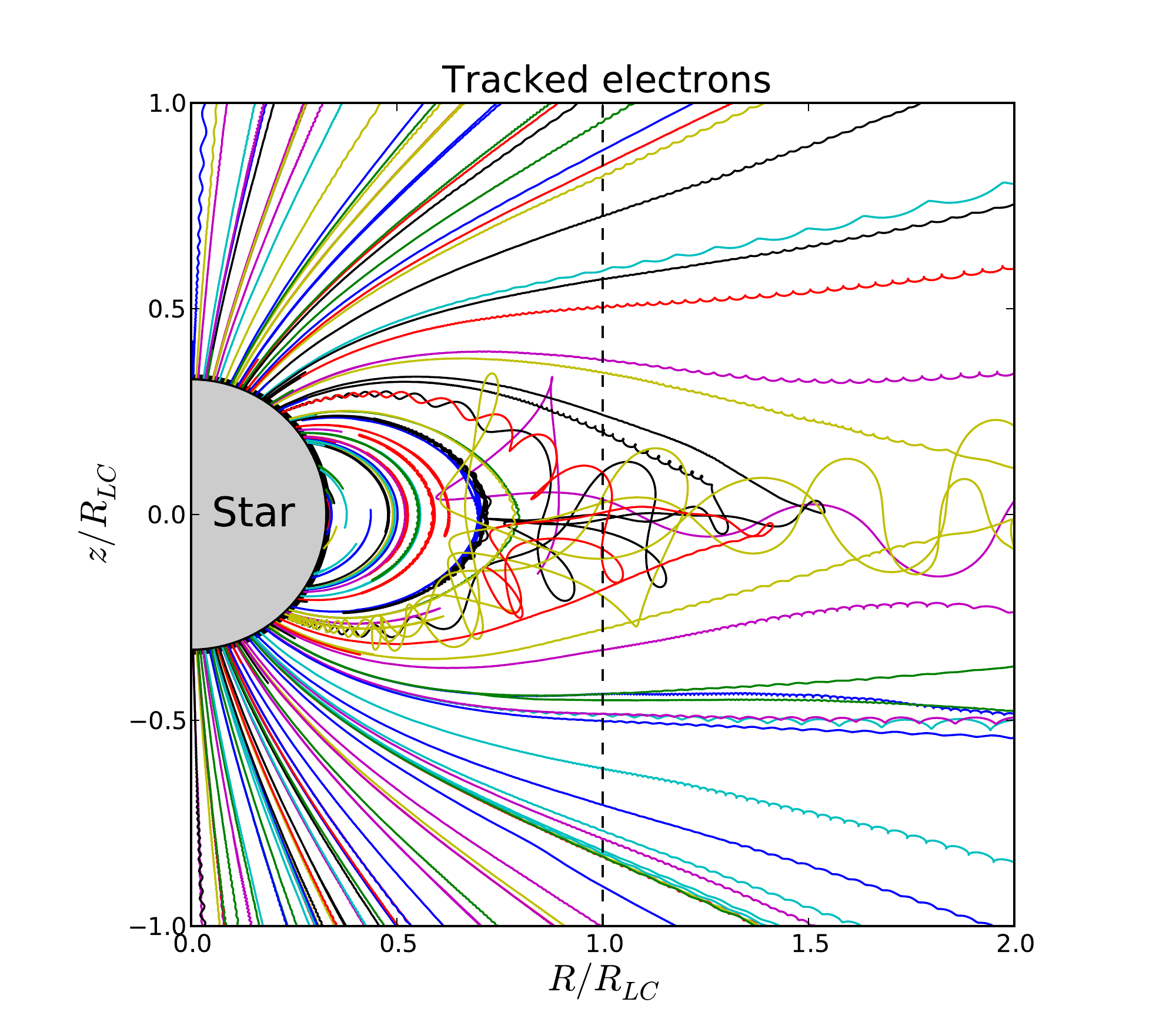}
\includegraphics[width=9cm]{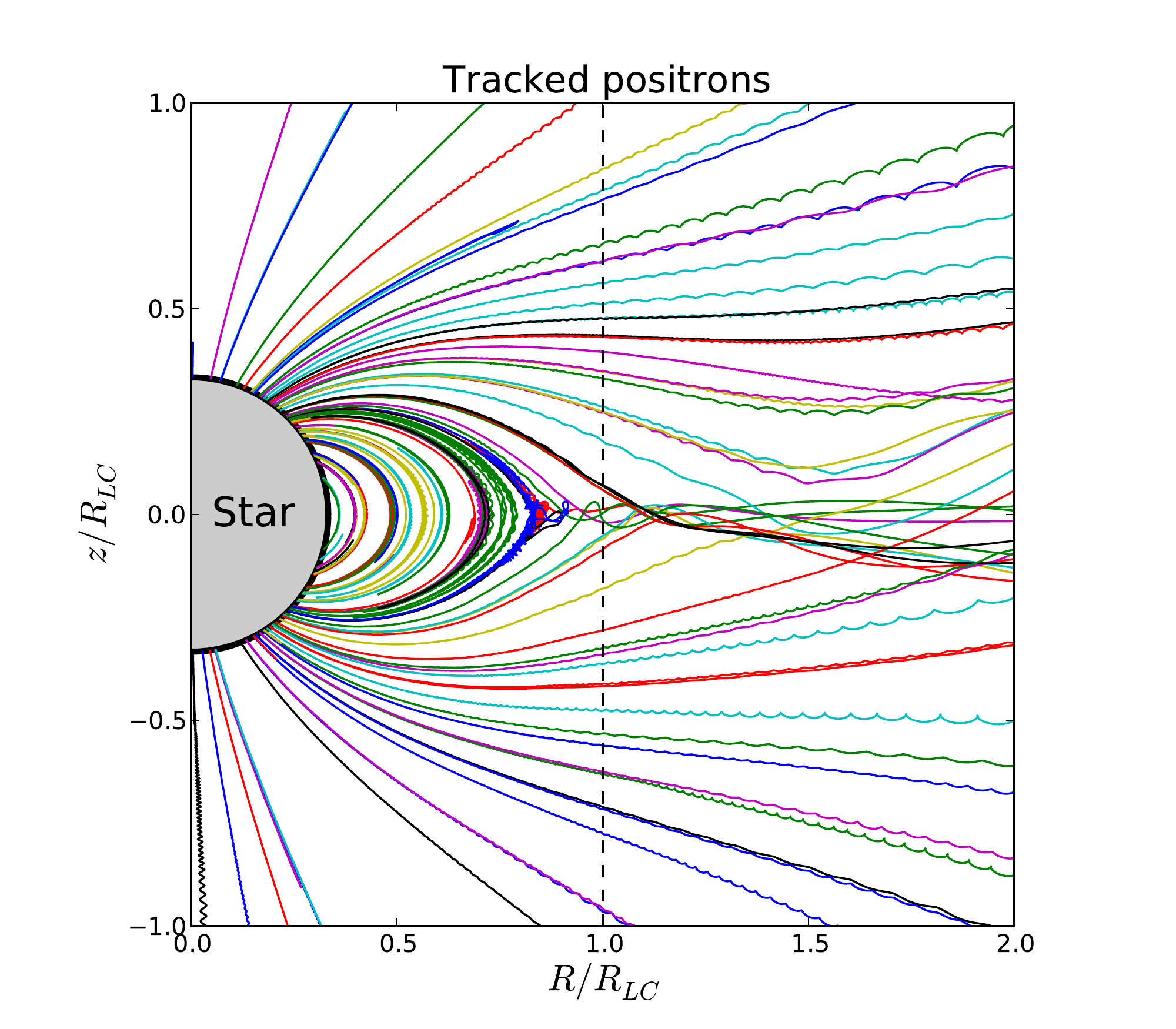}
\caption{Trajectories of a sample of 157 electrons (top) and 121 positrons (bottom) followed from the time of their injection at $t=2.25P$ at the surface of the star, for about 2 rotation periods, extracted from the run with $f_{\rm inj}=1$ and $\sigma_{\star}=5000$. The vertical dashed line indicates the light-cylinder radius. Magnetic fields lines are omitted for clarity.}
\label{fig_all_particles}
\end{figure}

\begin{figure}
\centering
\includegraphics[width=9cm]{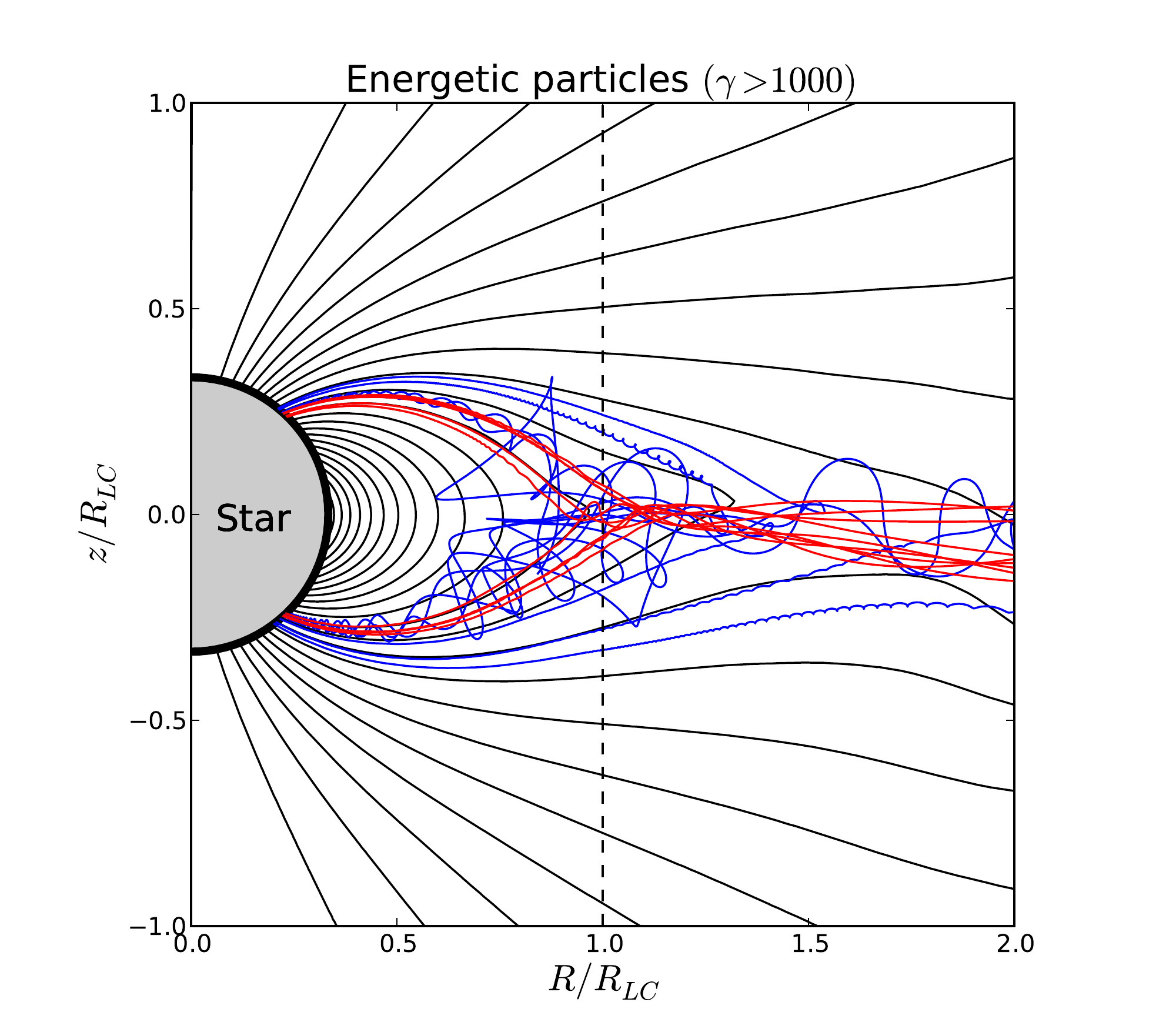}
\caption{Trajectories of all the particles (electrons in blue, positrons in red) that reach a Lorentz factor $\gamma>1000$ at least once in their history, from the sample shown in Fig.~\ref{fig_all_particles}. The black lines are magnetic field lines at the time when the particles were injected, i.e., at $t=2.25P$.}
\label{fig_HE_particles}
\end{figure}

Ultimately, the best way to understand how and where particles are accelerated is to look at the time history of individual particles. Fig.~\ref{fig_all_particles} shows the trajectories of $157$ electrons (top) and $121$ positrons (bottom) randomly selected from the newly injected pairs within $r=r_{\star}$ and $r=r_{\star}+\Delta r$, at $t/P=2.25$ for $f_{\rm inj}=1$ and $\sigma_{\star}=5000$. In this figure, one can distinguish the three regions discussed above: the closed and the open field line zones, and the current sheets. In the closed zone, the particles move along the same field line (they are well magnetized) from the star's surface where they are injected, to the other end of the field line back to the star, so that on average the number of particles in the closed zone is nearly constant. However, some particles are trapped by magnetic mirrors, i.e., they are reflected in regions of strong fields back to low field regions, and so on (like in the Van Allen belts in the Earth's magnetosphere). This effect may not occur in real pulsars because the particles are thought to have negligible transverse momenta. Along the open field lines, the particles are simply drifting outward along the quasi-radial field lines, and form the pulsar wind. At the poles, we note that some positrons are pushed back towards the star very soon after their injection (these particles are not shown here). This is due to the surface electric field that sorts the particles to maintain a net negative charge, and the required electric current, at the poles. Finally, the current sheets are identified by the particles moving along the boundary between the open and closed field line regions. Once these particles leave the separatrices, they are trapped along the equatorial current sheet where they follow relativistic Speiser orbits \citep{1965JGR....70.4219S, 2007A&A...472..219C, 2011ApJ...737L..40U, 2012ApJ...746..148C}.

\begin{figure*}
\centering
\includegraphics[width=7.5cm]{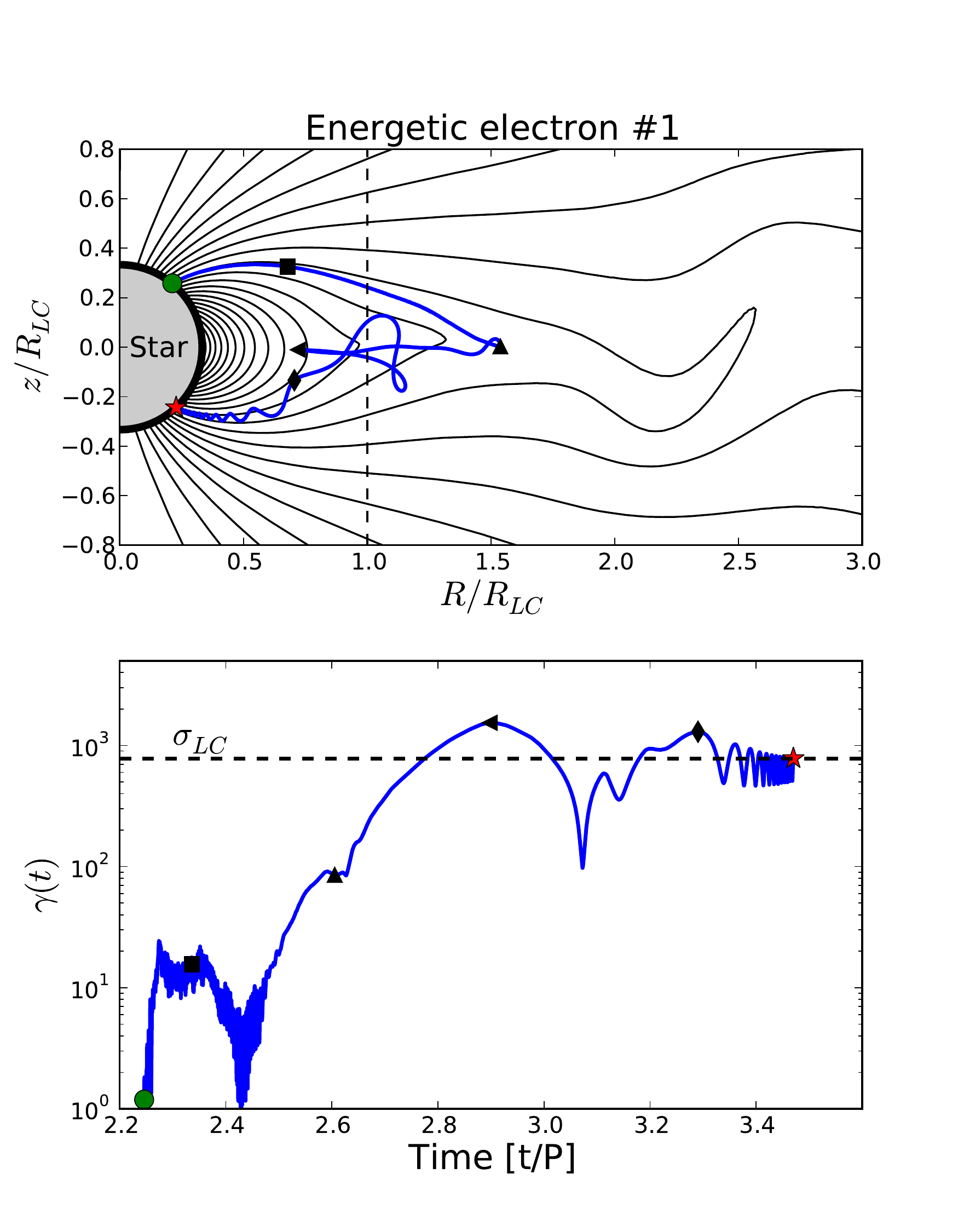}
\includegraphics[width=7.5cm]{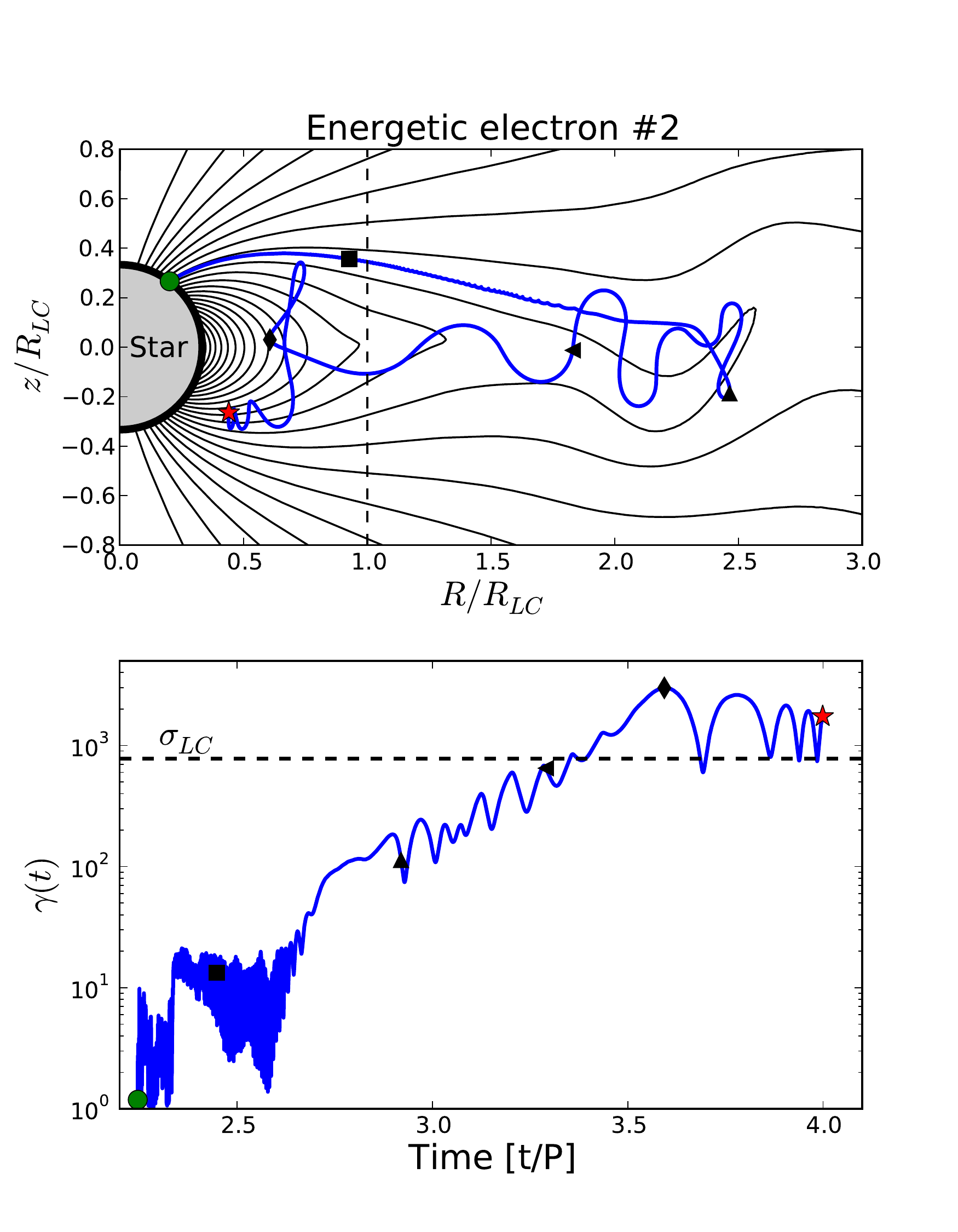}
\caption{Two representative trajectories of high-energy electrons from Fig.~\ref{fig_HE_particles} whose Lorentz factor reach at least $\gamma=1000$ during its whole time history (top panels). The magnetic field lines are shown at the time when the particles were injected ($t=2.25P$). The green disk shows the starting point of the particle, while the red star shows the location of the particle at the end of the simulation (top right) or once the particle returned back into the star (top left). The other symbols along the particle trajectories indicate intermediate times, which are carried over to the Lorentz factor time history curves on the bottom panels. The dashed horizontal line shows $\sigma_{\rm LC}=780$.}
\label{fig_electrons}
\end{figure*}

\begin{figure*}
\centering
\includegraphics[width=7.5cm]{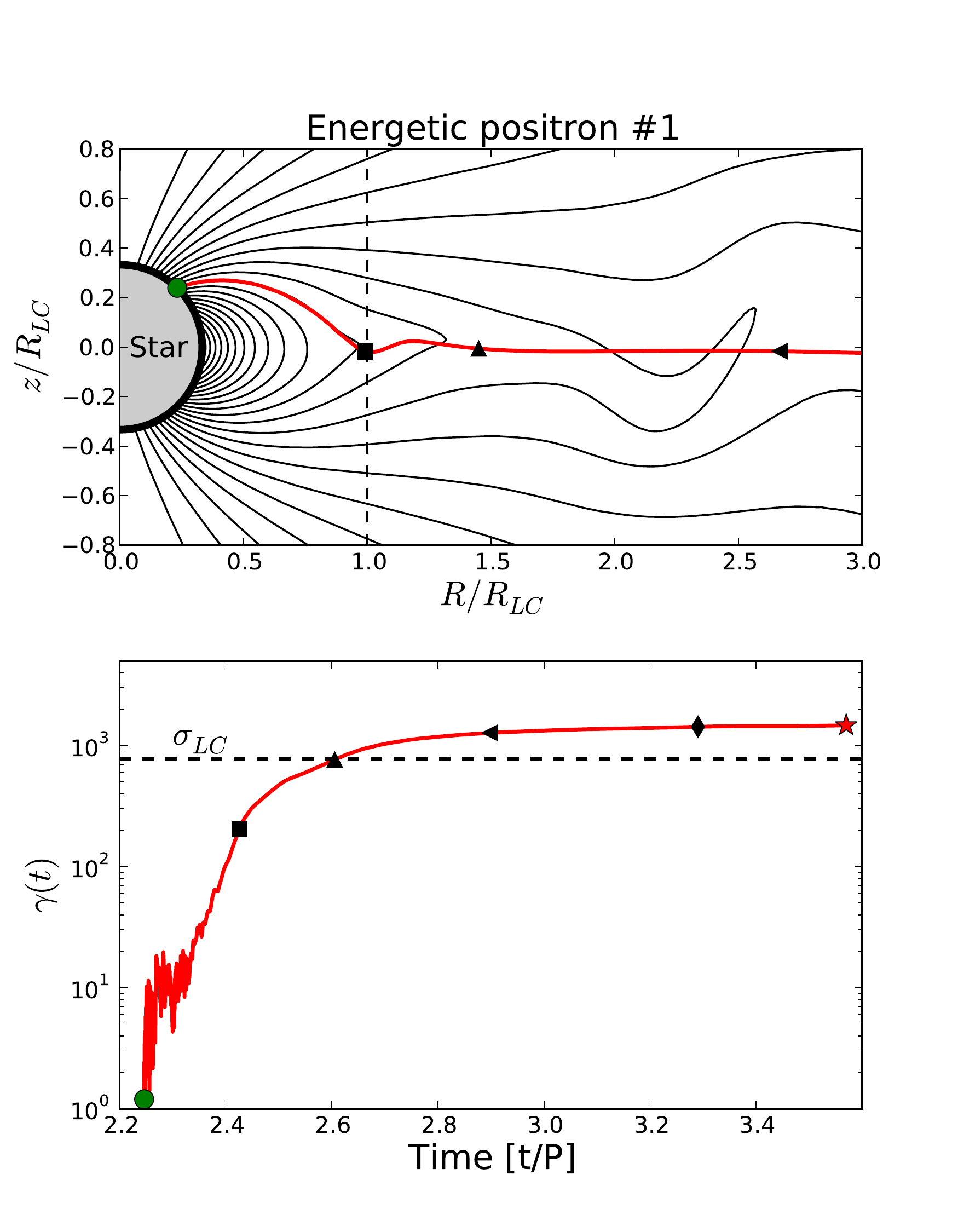}
\includegraphics[width=7.5cm]{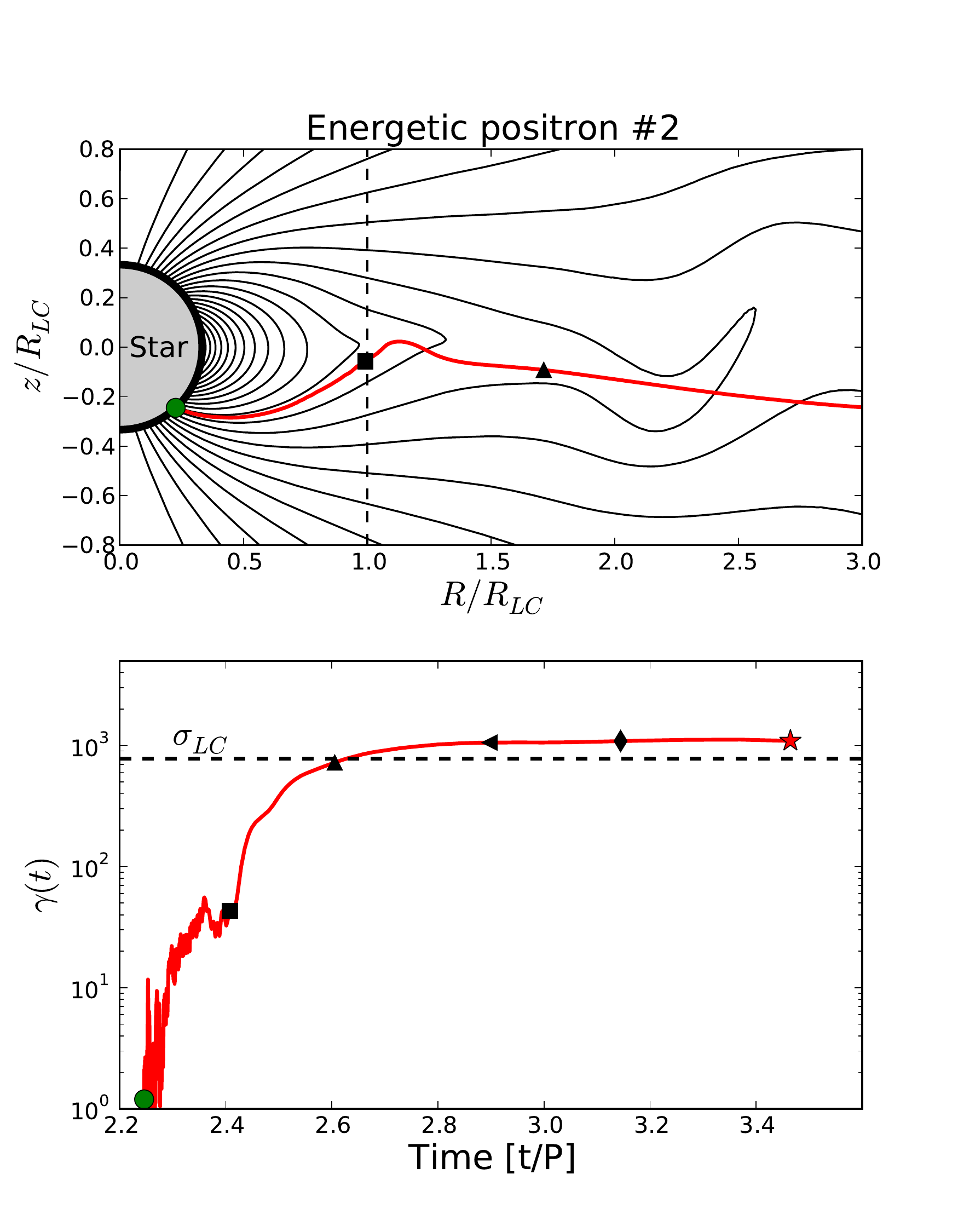}
\caption{Same as in Fig.~\ref{fig_electrons}, but for 2 representative high-energy positrons. Note that only part of the trajectories are shown here. Beyond $R=3R_{\rm LC}$, the high-energy positrons continue their path in a straight line until they leave the box through the outer boundary.}
\label{fig_positrons}
\end{figure*}

We identify the regions of intense particle acceleration by selecting the particles according to their energy only. Fig.~\ref{fig_HE_particles} shows the trajectories of all the energetic particles from the sample (electrons in blue, positrons in red) whose Lorentz factor reach at least $\gamma=1000$ during their history. These high-energy electrons represent about $2\%$ of the electrons and about $40\%$ of the total energy carried away by the electrons (see Fig.~\ref{fig_spectra}). Energetic positrons represent about $10\%$ of all the positrons, carrying about $50\%$ of the total positron energy. Drawn together, these orbits are exclusively confined to the current sheet regions (both the separatrices and the equatorial sheets, as shown in Fig.~\ref{fig_temperature}). More specifically, all the high-energy particles originate from a narrow range of magnetic footpoints on the surface of the star, spanning from about $\theta\sim 42^{\rm o}$-$44^{\rm o}$ for the positrons, and $\theta\sim 37^{\rm o}$-$40^{\rm o}$ for the electrons (note that these angles depend on the $r_{\star}/R_{\rm LC}$ ratio).

The particles get most of their energy once they reach the equatorial current sheet. Figs.~\ref{fig_electrons}, \ref{fig_positrons} show representative trajectories of energetic electrons and positrons taken from the sample, as well as the time evolution of their Lorentz factors. For positrons, after a first energy gain from $\gamma\approx 1$ to $\gamma\gtrsim 10$ close to the star, the bulk of the acceleration occurs at the Y-point over a distance of order the light-cylinder radius (see Fig.~\ref{fig_positrons}). Thereafter, their energy increases little with distance. In contrast, the high-energy electrons enter the current sheet further away, typically at $r\gtrsim 1$-$2R_{\rm LC}$ and precipitate back towards the star, counter-streaming against the outflowing positrons, in agreement with the two-fluid picture drawn in the previous section (Fig.~\ref{fig_bulk}, top panel). As the electrons precipitate, their energy increases almost linearly until they reach the Y-point. At this point, they bounce on the closed zone and flow back towards the star through the separatrices (see Fig.~\ref{fig_electrons}). The counter-streaming of electrons and positrons is maintained throughout the simulation by a non-zero radial electric field in the equatorial current sheet (see also \citealt{2007A&A...472..219C}). Electron orbits in the sheet also differ from positron orbits because the energetic electrons are moving against the radial $\mathbf{E}\times\mathbf{B}$ drift while the positrons are moving with it\footnote{This effect was brought to our attention by the referee Ioannis Contopoulos.}. This difference is exaggerated by the large amplitude of the electron orbits with respect to the layer mid-plane. Presumably, these excursions would rapidly disappear in real pulsars because the perpendicular momentum of the particle would be efficiently radiated away. In addition, the thickness of the current sheet would decrease in reaction to the loss of pressure support in the presence of radiative losses \citep{2011PhPl...18d2105U, 2013ApJ...770..147C}. The effect of radiative colling on the pulsar magnetosphere will be explored elsewhere.

\section{Conclusions}\label{conclusions}

We have solved for the structure of the aligned pulsar magnetosphere as a function of the plasma injection rate and magnetization, using high-resolution 2D axisymmetric PIC simulations. Fresh electron-positron pairs are injected at a constant rate with a non-zero velocity from the surface of the star to mimic pair production. If the plasma supply is low, the magnetosphere is highly charge-separated with electrons concentrated at the poles and positrons mostly in the equatorial regions. The pulsar luminosity is smaller than the expected force-free spin-down power of an aligned rotator, and presents a high dissipation rate ($\gtrsim 40\%$) of the Poynting flux beyond the light cylinder. This solution resembles the model recently proposed by \citet{2014ApJ...781...46C}, in which there are no separatrices or Y-point and the current layer is electrostatically supported. In the extreme regime where there is no pair production, our solution collapses to the static disk-dome solution of ``dead'' pulsars.

In contrast, we recover a nearly force-free solution of the magnetosphere in the high-multiplicity regime, with the expected spin-down power of an aligned pulsar. A strong current layer forms self-consistently beyond the light cylinder and along the equator, which supports open, quasi-monopolar, magnetic field lines. We find that about $30\%$ of the outgoing Poynting flux is dissipated in the current layer, mostly in the vicinity of the Y-point \citep{2014ApJ...785L..33P}. These results imply that the pulsar magnetosphere is highly sensitive to dissipation, consistent with earlier force-free simulations, having resistivity effectively confined to the equatorial current sheet, in which more than 20\% of the spin-down power was dissipated within ten light-cylinder radii \citep{2012MNRAS.423.1416P}.

This dissipated energy is efficiently transferred to particles in the current sheet. The simulations show that the mean Lorentz factor of the energetic particles is given by the upstream magnetization parameter at the light cylinder, i.e., $\langle\gamma\rangle\approx\sigma_{LC}$ \citep{2014ApJ...785L..33P}. We found that the high-energy particles follow one of two particular trajectories in the magnetosphere. They all originate from the boundary layer between the closed and open magnetic field line regions, with a slight offset between electrons and positrons. High-energy positrons stream outward along the separatrix with little change of energy until they reach the Y-point. At this point, there are linearly accelerated in the equatorial sheet by the reconnection electric field, over a distance of order the light-cylinder radius. Energetic electrons only enter the layer further downstream and precipitate back towards the star. Electrons are energized on their way to the Y-point where their acceleration stops abruptly; they then flow back onto the star along the separatrix. This mechanism naturally leads to an excess of energetic positrons flying away from the pulsar. In another publication, we will explore if this result can be related to the contribution of pulsars to the rising positron fraction measured by PAMELA \citep{2009Natur.458..607A}. 

Our results suggest that the current sheet, and the separatrix layers, should be intense sources of high-energy radiation in pulsars \citep{1996A&A...311..172L, 2002A&A...388L..29K, 2010ApJ...715.1282B, 2012MNRAS.424.2023P, 2013A&A...550A.101A, 2014ApJ...780....3U}. The precipitation of the energetic electrons onto the star may be an additional source of radiation, an ``auroral''-like emission \citep{2012SSRv..173..341A}. The particles in the wind (outside the layer) follow the $\mathbf{E}\times\mathbf{B}$ drift velocity. These particles experience no radiative losses and little radiation is expected from this region, unless they upscatter background radiation via inverse-Compton scattering \citep{2000APh....12..335B, 2000MNRAS.313..504B, 2008A&A...488...37C, 2012Natur.482..507A}. Our simulations ignore synchrotron or curvature radiation cooling which should be important in young pulsars, although we anticipate that the overall structure of the magnetosphere and the acceleration mechanism reported here should be preserved. The other important limitation of this work is the simplistic particle injection. The next logical step towards the understanding of the high-energy emission in pulsars is to perform 3D PIC simulations of the oblique rotator, and to include the physics of pair production in the magnetosphere \citep{2014arXiv1412.0673P}.

\section*{Acknowledgments}
We thank J. Arons, A. Beloborodov, A. Chen, D. Caprioli, L. Sironi, A. Tchekhovskoy, and D.~A. Uzdensky for discussions, and the referee Ioannis Contopoulos for his valuable comments on the manuscript. BC acknowledges support from the Lyman Spitzer Jr. Fellowship awarded by the Department of Astrophysical Sciences at Princeton University. This work was partially suppored by the Max-Planck/Princeton Center for Plasma Physics, by NASA grants NNX12AD01G and NNX13AO80G, and by the Simons Foundation (grant 291817 to AS). Computing resources were provided by the PICSciE-OIT TIGRESS High Performance Computing Center and Visualization Laboratory at Princeton University.

\bibliographystyle{aa}
\bibliography{pulsar2D}

\onecolumn

\appendix

\section{Integration of Maxwell's equations on the 2D spherical Yee grid}\label{app_maxwell}

\begin{figure}
\centering
\includegraphics[width=6cm]{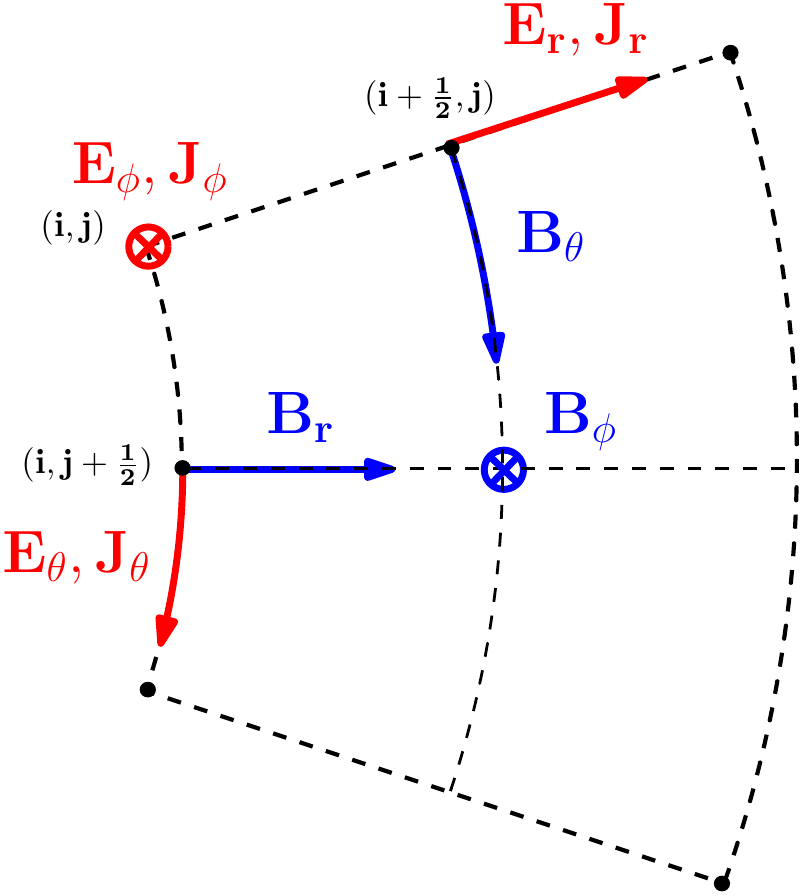}
\caption{Two-dimensional spherical Yee-lattice adopted in {\tt Zeltron}.}
\label{fig_yee}
\end{figure}

In this appendix, we derive the expressions of the differential operators $\mathbf{\nabla}\times\mathbf{E},~\mathbf{\nabla}\times\mathbf{B}$, $\mathbf{\nabla}\cdot\mathbf{E}$ and $\mathbf{\nabla}^2$ integrated over a 2D axisymmetric cell defined in spherical coordinates (so that any derivative in the azimuthal direction vanishes, $\partial/\partial\phi=0$), as implemented in {\tt Zeltron} to solve Maxwell's equations. Each cell is labeled by the integers (i,j), i for the radial direction, and j for the $\theta$-direction. The fields are defined on the Yee lattice which is staggered by half a grid cell in both spatial directions, as shown in Fig.~\ref{fig_yee}.

$\mathbf{\nabla}\times \mathbf{E}$ and $\mathbf{\nabla}\times \mathbf{B}$ can easily be integrated using Stokes' theorem on one cell, i.e.,
\begin{equation}
\iint_{\mathcal{S}}(\mathbf{\nabla}\times\mathbf{E})\cdot d{\bf \mathcal{S}}=\oint_{\mathcal C}\mathbf{E}\cdot d\bf{\mathcal{C}},
\label{stokes}
\end{equation}
where $\bf{\mathcal{S}}$ is the surface vector pointing away from the cell, and $\bf{\mathcal{C}}$ is the contour vector circulating around the cell. Applying Eq.~(\ref{stokes}) in each direction gives
\begin{eqnarray}
\left(\mathbf{\nabla}\times\mathbf{E}\right)_{r_{\rm i,j+\frac{1}{2}}} &=& \frac{\sin\theta_{\rm j+1}E_{\phi_{\rm i,j+1}}-\sin\theta_{\rm j}E_{\phi_{\rm i,j}}}{r_{\rm i}\Delta\mu_{\rm j+\frac{1}{2}}}\\
\left(\mathbf{\nabla}\times\mathbf{E}\right)_{\rm{\theta_{i+\frac{1}{2},j}}} &=& -\frac{2\left(r_{\rm i+1}E_{\phi_{\rm i+1,j}}-r_{\rm i}E_{\phi_{\rm i,j}}\right)}{\Delta r^2_{\rm i+\frac{1}{2}}}\\
\left(\mathbf{\nabla}\times\mathbf{E}\right)_{\phi_{\rm i+\frac{1}{2},j+\frac{1}{2}}} &=& -\frac{2\Delta r_{\rm i+\frac{1}{2}}\left(E_{r_{\rm i+\frac{1}{2},j+1}}-
E_{r_{i+\frac{1}{2},j}}\right)}{\Delta r^2_{\rm i+\frac{1}{2}}\Delta\theta}+\frac{2\left(r_{\rm i+1}E_{\theta_{\rm i+1,j+\frac{1}{2}}}-r_{\rm i}E_{\theta_{\rm i,j+\frac{1}{2}}}\right)}{\Delta r^2_{\rm i+\frac{1}{2}}},
\label{rotE}
\end{eqnarray}
where $\Delta\mu_{\rm j+\frac{1}{2}}=\left(\cos\theta_{\rm j}-\cos\theta_{\rm j+1}\right)$, $\Delta r_{\rm i+\frac{1}{2}}=\left(r_{\rm i+1}-r_{\rm i}\right)$ and $\Delta r^2_{\rm i+\frac{1}{2}}=\left(r^2_{\rm i+1}-r^2_{\rm i}\right)$. Similarly, integrating $\mathbf{\nabla}\times\mathbf{B}$ on the Yee lattice yields
\begin{eqnarray}
\left(\mathbf{\nabla}\times\mathbf{B}\right)_{r_{\rm i+\frac{1}{2},j}} &=& \frac{\sin\theta_{\rm j+\frac{1}{2}}B_{\phi_{\rm i+\frac{1}{2},j+\frac{1}{2}}}-\sin\theta_{\rm j-\frac{1}{2}}B_{\phi_{\rm i+\frac{1}{2},j-\frac{1}{2}}}}{r_{i+\frac{1}{2}}\Delta\mu_{\rm j}}\\
\left(\mathbf{\nabla}\times\mathbf{B}\right)_{\theta_{\rm i,j+\frac{1}{2}}} &=& -\frac{2\left(r_{\rm i+\frac{1}{2}}B_{\phi_{\rm i+\frac{1}{2},j+\frac{1}{2}}}-r_{\rm i-\frac{1}{2}}B_{\phi_{\rm i-\frac{1}{2},j+\frac{1}{2}}}\right)}{\Delta r^2_{\rm i}}\\
\left(\mathbf{\nabla}\times\mathbf{B}\right)_{\phi_{\rm i,j}} &=& -\frac{2\Delta r_{\rm i}\left(B_{r_{\rm i,j+\frac{1}{2}}}-
B_{r_{\rm i,j-\frac{1}{2}}}\right)}{\Delta r^2_{\rm i}\Delta\theta}+\frac{2\left(r_{\rm i+\frac{1}{2}}B_{\theta_{\rm i+\frac{1}{2},j}}-r_{\rm i-\frac{1}{2}}B_{\theta_{\rm i-\frac{1}{2},j}}\right)}{\Delta r^2_{\rm i}},
\label{rotB}
\end{eqnarray}
with $\Delta\mu_{\rm j}=\left(\cos\theta_{\rm j-\frac{1}{2}}-\cos\theta_{\rm j+\frac{1}{2}}\right)$, $\Delta r_{\rm i}=\left(r_{\rm i+\frac{1}{2}}-r_{\rm i-\frac{1}{2}}\right)$ and $\Delta r^2_{\rm i}=\left(r^2_{\rm i+\frac{1}{2}}-r^2_{\rm i-\frac{1}{2}}\right)$. 

To integrate $\mathbf{\nabla}\cdot \mathbf{E}$, we make use of Gauss' theorem
\begin{equation}
\iiint_{\mathcal{V}} \left(\mathbf{\nabla\cdot\ E}\right)d\mathcal{V}=\oiint_{\mathcal{S}}\mathbf{E}\cdot d\mathbf{S}
\label{gauss}
\end{equation}
where $\bf{\mathcal{V}}$ is the volume of the cell, and $\bf{\mathcal{S}}$ is the surface vector pointing away from the cell. It gives
\begin{eqnarray}
\left(\mathbf{\nabla}\cdot\mathbf{E}\right)_{\rm i,j}=\frac{3}{\Delta r^3_{\rm i}}\left(r^2_{\rm i+\frac{1}{2}}E_{r_{\rm i+\frac{1}{2},j}}-r^2_{\rm i-\frac{1}{2}}E_{r_{\rm i-\frac{1}{2},j}}\right)+\frac{3\Delta r^2_{\rm i}}{2\Delta r^3_{\rm i}\Delta\mu_{\rm j}}\left(\sin\theta_{\rm j+\frac{1}{2}}E_{\theta_{\rm i,j+\frac{1}{2}}}-\sin\theta_{\rm j-\frac{1}{2}}E_{\theta_{\rm i,j-\frac{1}{2}}}\right),
\label{divE}
\end{eqnarray}
with $\Delta r^3_{\rm i}=\left(r^3_{\rm i+\frac{1}{2}}-r^3_{\rm i-\frac{1}{2}}\right)$.

These expressions are particularly interesting for this problem because there is no division by zero on the axis (in contrast to the differential expressions). With these expressions at hand and using Maxwell's equations, the fields are given by
\begin{eqnarray}
\mathbf{E}^{\rm n+1}&=&\mathbf{E}^{\rm n}+\Delta t\left[c\left(\mathbf{\nabla}\times\mathbf{B}\right)^{\rm n+\frac{1}{2}}-4\pi\mathbf{J}^{\rm n+\frac{1}{2}}\right]\\
\mathbf{B}^{\rm n+\frac{1}{2}}&=&\mathbf{B}^{\rm n-\frac{1}{2}}-c\Delta t\left(\mathbf{\nabla}\times\mathbf{E}\right)^{\rm n},
\end{eqnarray}
at the time step ${\rm n+1}$ for the electric field and ${\rm n+\frac{1}{2}}$ for the magnetic field. The time step $\Delta t$ is dictated by the Courant-Friedrichs-Lewy condition which, in spherical coordinates, gives 
\begin{equation}
\Delta t\leq \frac{1}{c\sqrt{1/\left({\rm Min}\left(\Delta r\right)\right)^2+1/\left({\rm Min}\left(r\Delta\theta\right)\right)^2}}.
\end{equation}
Because {\tt Zeltron} does not conserve electric charge exactly, the electric field must be corrected by solving Poisson's equation. If $\mathbf{E}$ is the electric field given by the code, it must be corrected by $\delta\mathbf{E}$ such that ${\mathbf\nabla}\cdot\left(\mathbf{E}+\delta\mathbf{E}\right)=4\pi\rho$, where $\rho$ is the charge density. Poisson's equation yields
\begin{equation}
\mathbf{\nabla}\cdot\mathbf{\nabla}\left(\delta\Phi\right)=-\left(4\pi\rho-\mathbf{\nabla}\cdot\mathbf{E}\right),
\label{poisson}
\end{equation}
where $\delta\mathbf{E}=-\mathbf{\nabla}\left(\delta\Phi\right)$. Applying Eq.~(\ref{divE}) in Eq.~(\ref{poisson}), one finds
\begin{eqnarray}
\delta\Phi_{\rm i,j}\left[\frac{3}{\Delta r^3_{\rm i}}\left(\frac{r^2_{\rm i+\frac{1}{2}}}{\Delta r_{\rm i+\frac{1}{2}}}+\frac{r^2_{\rm i-\frac{1}{2}}}{\Delta r_{\rm i-\frac{1}{2}}}\right)+\frac{3\Delta r^2_{\rm i}}{2\Delta r^3_{\rm i}\Delta\mu_{\rm j} r_{\rm i} \Delta\theta}\left(\sin\theta_{\rm j+\frac{1}{2}}+\sin\theta_{\rm j-\frac{1}{2}}\right)\right]=\left[4\pi\rho_{\rm i,j}-\left(\mathbf{\nabla}\cdot\mathbf{E}\right)_{\rm i,j}\right]+\nonumber \\
\frac{3}{\Delta r^3_{\rm i}}\left[\frac{r^2_{\rm i+\frac{1}{2}}}{\Delta r_{\rm i+\frac{1}{2}}}\delta\Phi_{\rm i+1,j}+\frac{r^2_{\rm i-\frac{1}{2}}}{\Delta r_{\rm i-\frac{1}{2}}}\delta\Phi_{\rm i-1,j}\right]+{\rm \frac{3\Delta r^2_{\rm i}}{2\Delta r^3_{\rm i}\Delta\mu_{\rm j} r_{\rm i} \Delta\theta}\left(\sin\theta_{\rm j+\frac{1}{2}}\delta\Phi_{\rm i,j+1}+\sin\theta_{\rm j-\frac{1}{2}}\delta\Phi_{\rm i,j-1}\right)}.
\label{phi}
\end{eqnarray}
After many iterations (typically about 500), we obtain from Eq.~(\ref{phi}) $\delta\Phi_{\rm i,j}$, and hence the correct electric field. This procedure is repeated every $25$ timesteps in the simulations presented here. At the boundaries of the domain, we apply a zero gradient condition, such that $\partial(\delta\Phi)/\partial r=0$, $\partial(\delta\Phi)/\partial\theta=0$.

\begin{figure}
\centering
\includegraphics[width=8cm]{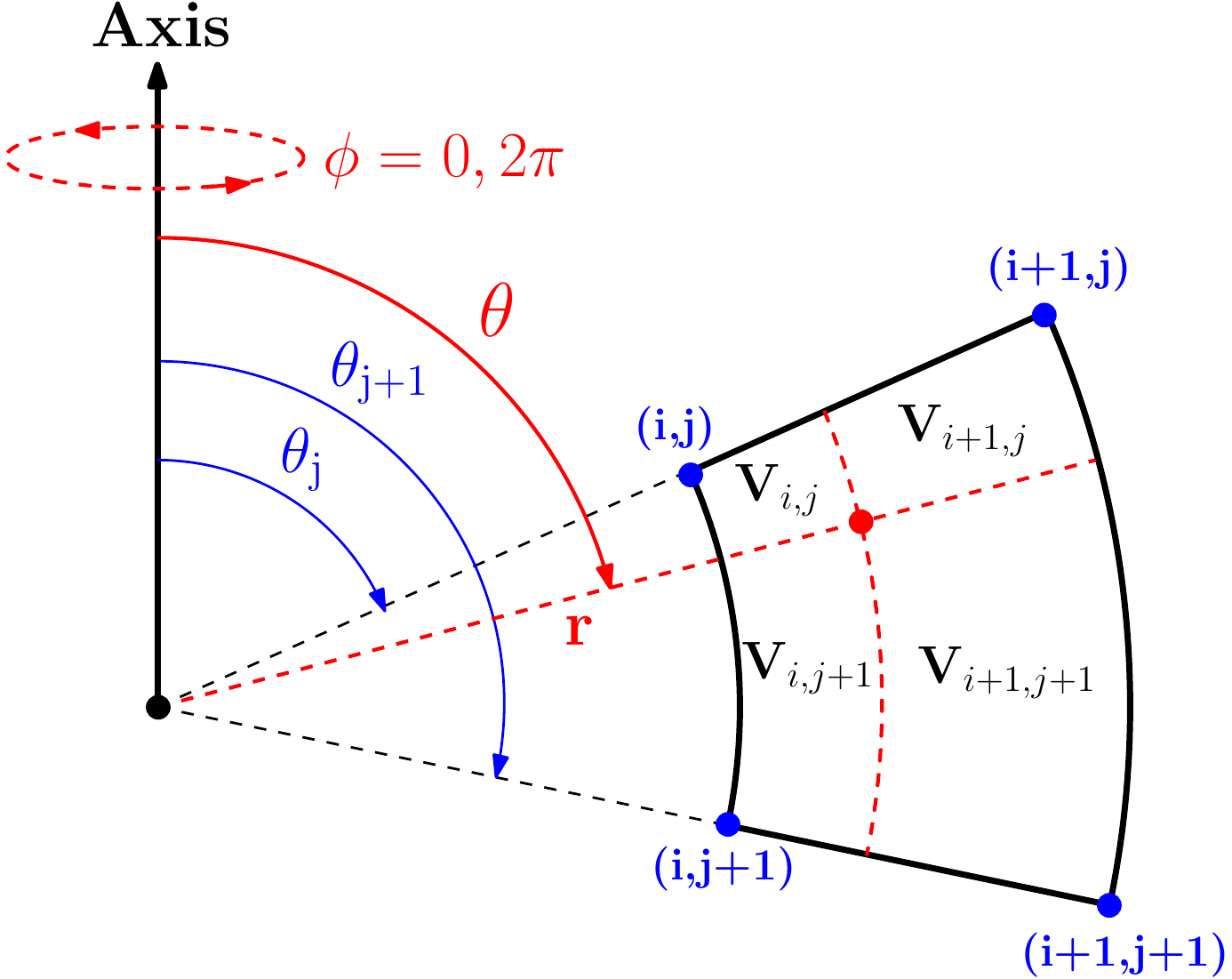}
\caption{This diagram represents the geometry of a single cell in a 2D axisymmetric spherical mesh, as well as the volumes involved in the interpolation scheme utilized in {\tt Zeltron} (see text for details).}
\label{fig_interpol}
\end{figure}

\section{Interpolation scheme}\label{app_deposit}

This section presents the 2D linear interpolation scheme used in the {\tt Zeltron} code to deposit the charge and current carried by the particles on the spherical grid. Consider a particle of charge $Q$ located at $(r,\theta)$, where $r_{\rm i}<r<r_{\rm i+1}$ and $\theta_{\rm j}<\theta<\theta_{\rm j+1}$. Then, the amount of charge deposited on the closest grid nodes is given by the ratio of the volumes shown in Fig.~\ref{fig_interpol}, such that
\begin{eqnarray}
Q_{\rm i,j}&=&\frac{V_{\rm i+1,j+1}}{V_{\rm tot}}Q=\left(1-f_{\rm r}\right)\left(1-f_{\theta}\right)Q\\
Q_{\rm i+1,j}&=&\frac{V_{\rm i,j+1}}{V_{\rm tot}}Q=f_{\rm r}\left(1-f_{\theta}\right)Q\\
Q_{\rm i,j+1}&=&\frac{V_{\rm i+1,j}}{V_{\rm tot}}Q=\left(1-f_{\rm r}\right)f_{\theta}Q\\
Q_{\rm i+1,j+1}&=&\frac{V_{\rm i,j}}{V_{\rm tot}}Q=f_{\rm r}f_{\theta}Q,
\end{eqnarray}
where $V_{\rm tot}=\frac{2\pi}{3}\left(r^3_{\rm i+1}-r^3_{\rm i}\right)\left(\mu_{\rm j}-\mu_{\rm j+1}\right)$ is the total volume of the cell, and
\begin{eqnarray}
f_{\rm r}&=&\frac{r^3-r^3_{\rm i}}{r^3_{\rm i+1}-r^3_{\rm i}}\\
f_{\theta}&=&\frac{\cos\theta_{\rm j}-\cos\theta}{\cos\theta_{\rm j}-\cos\theta_{\rm j+1}}.
\label{pq}
\end{eqnarray}
The same numerical scheme is used in the Boris push to interpolate the fields from the spherical grid to the location of the particles.

\label{lastpage}

\end{document}